\documentclass[letterpaper,10pt, onecolumn, twoside, nosectionnumbsers]{classes/LTJournalArticle}

\def\doctitle{Infrared-absorption magnetometry in a diamond nanocavity: demonstration and limits at the nanoscale}

\def\authorOne{Nicholas J. Sorensen}
\def\authorNine{Hadi Zadeh-Haghighi}
\def\authorTwo{Chris Coutts}
\def\authorThree{Sean McNaney}
\def\authorThreeA{Matteo Feder}
\def\authorFour{Elham Zohari}
\def\authorFive{Sigurd Fl{\aa}gan}
\def\authorSix{Vinaya~K.~Kavatamane}
\def\authorSeven{Paul E. Barclay}
\def\authorEight{Christoph Simon}
\def\addressOne{
    Institute for Quantum Science and Technology, University of Calgary, Calgary, Alberta, T2N 1N4, Canada
    }
\def\addressTwo{Department of Physics, University of Alberta, Edmonton, Alberta, T6G 2R3, Canada}
\def\addressThree{National Research Council of Canada, Quantum and Nanotechnology Research Centre, Edmonton, Alberta, T6G 2M9, Canada}

\def\emailContact{nicholas.sorensen@ucalgary.ca}

\def\abstractText{
Diamond quantum sensors promise both high spatial resolution and high sensitivity; however, conventional photoluminescence-based spin readout can be difficult to integrate efficiently.
Infrared absorption by nitrogen-vacancy (NV) centres in diamond provides a promising alternative, although the sensitivity of compact absorption-based sensors is limited by the short optical interaction length between the probe field and the NV ensemble.
To overcome this limitation, we demonstrate, for the first time, singlet-absorption magnetometry using spin-dense diamond nanocavities.
By probing cavity modes near the 1042~nm NV singlet transition, we observe spin-dependent cavity transmission and perform cavity-enhanced magnetic field sensing.
We further develop and experimentally validate a model that accurately predicts the spin-dependent cavity response, describes the optimal operating conditions, and identifies the practical limitations of different optically detected magnetic resonance modalities.
Together, these results establish cavity-enhanced singlet absorption as a scalable readout mechanism for integrated diamond quantum sensors and provide a practical framework for engineering next-generation nanophotonic diamond quantum magnetometers.
}
\usepackage[utf8]{inputenc}
\usepackage[T1]{fontenc}

\usepackage{hyperref}

\usepackage{upgreek}
\usepackage{dblfloatfix}

\usepackage{newtxtext}
\usepackage{newtxmath}
\usepackage[nameinlink]{cleveref}
\usepackage{soul}
\usepackage{graphicx}%
\usepackage{multirow}%
\usepackage[dvipsnames]{xcolor}
\usepackage{textcomp}%
\usepackage{lineno}
\usepackage{mathtools}
\usepackage{natbib}
\usepackage{fix-cm} 
\usepackage{dcolumn}

\usepackage{orcidlink,breqn,enumitem,caption,subcaption,ulem,titlesec,mathtools,gensymb,float,setspace,multicol}

\Crefname{equation}{Eq.}{Eqs.}
\Crefname{figure}{Fig.}{Figs.}
\Crefname{table}{Tab.}{Tabs.}
\Crefname{section}{Appendix}{Appendices}

\AtBeginDocument{\typeout{HBAR = \meaning\hbar}}
\begin{document}
\title{\doctitle}

\author{\normalsize
\authorOne\orcidlink{0000-0002-4666-5791},\textsuperscript{1,*}
\authorFour\orcidlink{0000-0001-8344-4157} ,\textsuperscript{1,2,3}
\authorTwo\orcidlink{0009-0007-7119-9896},\textsuperscript{1} \\   \normalsize 
\authorThree,\textsuperscript{1}
\authorThreeA,\textsuperscript{1}
\authorNine\orcidlink{0000-0003-3380-9925},\textsuperscript{1}
\authorFive\orcidlink{0000-0003-0272-7601},\textsuperscript{1} \\ \normalsize
\authorSix\orcidlink{0000-0002-4287-9290},\textsuperscript{1}
\authorEight\orcidlink{0000-0003-0747-8800},\textsuperscript{1} and 
\authorSeven\orcidlink{0000-0002-9659-5883}\textsuperscript{1}}
\date{\small\textit{\textsuperscript{1}\addressOne  \\ 
\textsuperscript{2}\addressTwo \\ 
\textsuperscript{3}\addressThree \\
\textsuperscript{*}\emailContact}}

\twocolumn[
\maketitle
\begin{abstract}
    \abstractText
\end{abstract}
]

\section{Introduction}

\begin{figure}[b!]
    \centering
    \includegraphics[]{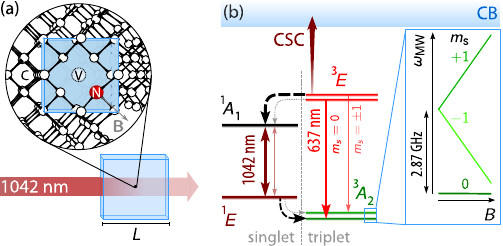}
    \caption{\textbf{ODMR using singlet absorption by the diamond NV centre.} {(a)} The NV centre consists of a substitutional nitrogen atom (N) adjacent to a vacancy (V) in the diamond carbon (C) lattice. 
    {(b)} The negatively charged NV centre has a ground-state spin triplet ($|m_{\text{s}} = 0, \pm1\rangle$) with Zeeman-split sublevels in a magnetic field $\mathbf{B}$.
    Optical excitation from $^3\!A_2\rightarrow\,^3\!E$ is followed by spin-dependent decay through either the triplet or singlet manifolds, enabling optical spin readout. 
    Upon optical excitation, NVs with $|m_{\text{s}} = \pm1 \rangle\,(|m_{\text{s}} = 0\rangle)$ produce less (more) triplet PL and more (less) singlet PL and are more (less) effective at absorbing an IR probe laser. Therefore,
    the spin state can be read by measuring the transmissivity of a 1042~nm probe field (a).
    Intense probe fields can induce charge-state conversion (CSC) from $^3\!E$~\cite{Hopper2016,Shandilya2025,Flagan2025,Itoi2026}. 
    This limits the probe intensity, thereby limiting the magnetic sensitivity. 
    $\omega_{\text{MW}}$ is the frequency of a microwave field used to control the ensemble spin-state.
    }
    \label{fig:fig1}
\end{figure}

Quantum sensors based on solid-state spin defects enable highly sensitive measurements of magnetic fields, temperature, strain, and electric fields~\cite{Rondin2014,degen2017quantum,Shandilya2022}.
Among spin defect sensing platforms, the negatively charged nitrogen-vacancy (NV) centre in diamond [\Cref{fig:fig1}~(a)] has emerged as a leader due to its long room-temperature spin-coherence times~\cite{Balasubramanian2008,Jarmola2012} and optical spin readout~\cite{Barry2020,Shandilya2021}.
Sensors based on single NV centres can achieve nanoscale spatial resolution~\cite{Balasubramanian2008,Maze2008,Taylor2008}, and ensemble-based sensors can achieve magnetic field sensitivities approaching $100\,\text{fT}\,\text{Hz}^{-1/2}$\,\cite{Barry2024}, enabling applications in materials science~\cite{Maletinsky2012, Fu2014, Glenn2017, Zhou2020, Hedrich2021, Scheidegger2022, Bopp2025} and biosensing~\cite{Steinert2013, Barry2016, Fescenko2019, Kuwahata2020, Niora2026, Pointner2026,Heinen2026}. 
Sensors that combine high sensitivity and high spatial resolution are desirable for many applications, and nanofabricated diamond photonic devices can provide robust and compact platforms for achieving this goal ~\cite{Graham2023,Sorensen2025,Katsumi2025}. 

Magnetic field sensing with NV centres is most commonly performed using optically detected magnetic resonance (ODMR) to
probe the magnetic-field dependence of the NV's spin-triplet electronic ground state $^3\!A_2$ [\Cref{fig:fig1}~(b)], whose Zeeman-split spin transitions can be coherently driven using resonant microwave fields~\cite{Barry2020}.
A spin-dependent intersystem crossing from the triplet to a longer-wavelength singlet manifold results in spin-dependent changes to visible triplet photoluminescence (PL) under off-resonant optical excitation with a green laser. This enables conventional ODMR~\cite{Barry2020} measurements of changes to the $^3\!A_2$ ground state spin populations driven by a microwave field resonant with their magnetically sensitive spin transitions.

Two alternative ODMR modalities can be implemented by harnessing optical properties of the singlet manifold: its emission at infrared (IR) wavelengths near and above the singlet's zero-phonon line (ZPL, $\lambda_{\text{ZPL}}\approx1042$~nm) generated by the $^1\!A_1 \rightarrow {}^1\!E$ transition, and its absorption near $\lambda_{\text{ZPL}}$ from the $^1\!E\rightarrow\,^1\!A_1$ transition~\cite{Marini2008,Acosta2010}.
The strength of both of these signals depends on the NV population in the singlet manifold resulting from the intersystem crossing.

Although both of these singlet processes are hindered by the transition's comparatively weak optical cross-section, singlet-absorption based ODMR offers several advantages stemming from its operation at a single measurement wavelength. Magnetic sensitivity of PL readout schemes, whether based on singlet or triplet emission, is fundamentally limited by the quantum efficiency~\cite{Radko2016,Janitz2020} and emission rate~\cite{Tetienne2012} of the relevant optical transitions.
In practice, finite collection efficiency further constrains performance, despite substantial efforts to improve photon extraction ~\cite{Babinec2010,Hadden2010,LeSage2012,Hedrich2020,Barry2024}.
In contrast, singlet-absorption sensors ordinarily measure the transmission of a $\lambda_{\text{s}} \approx \lambda_{\text{ZPL}}$ wavelength continuous-wave (cw) probe laser through diamond~\cite{Acosta2010,Schall2025,TayefehYounesi2025} [\Cref{fig:fig1}~(a)] or within a composite Fabry--Perot cavity~\cite{Dumeige2013,Jensen2014,Gottesman2024}. 
The corresponding photon-shot-noise-limited (PSNL) sensitivity $\eta$ is primarily determined by the detected probe power $P_{\text{det}}$, together with the ODMR contrast $C$ and linewidth $\nu$~\cite{Drau2011}:
\begin{align}
\eta = \beta \frac{\nu}{C\sqrt{P_{\text{det}}}}\,,
\label{eq:sensitivity}
\end{align}
where $\beta \equiv {4\sqrt{\hbar\omega_s}}/({3\sqrt{3}\gamma_e})$, $\hbar$ is the reduced Planck constant, $\gamma_e$ is the NV gyromagnetic ratio, and $\omega_{\text{s}}$ is the IR probe frequency.
Compared with conventional triplet-PL ODMR, singlet-absorption ODMR can achieve improved sensitivity through both larger $P_{\text{det}}$ and $C$~\cite{Chatzidrosos2017,Zadeh-Haghighi2024}.
In many configurations, the coherent probe field can be collected with near-unity efficiency, avoiding collection losses associated with triplet PL, while the absorption contrast benefits from the spin selectivity of intersystem crossing to the singlet manifold and its insensitivity to PL from neutrally charged NVs~\cite{Waldherr2011}.

One drawback of singlet-absorption ODMR is that the spin contrast depends on the length of the absorber $L$ and the spin-dependent change in NV absorption.
For small sensors, $C$ is correspondingly small; however, integrating the sensors within optical cavities can enhance the interaction between the optical field and the spin defects~\cite{Janitz2020}, improving $C$ and, consequently, the magnetic sensitivity~\cite{Dumeige2013,Zadeh-Haghighi2024}.
Beyond increasing the effective interaction length, cavity integration introduces several other benefits for singlet-absorption magnetometry. 
These include suppression of probe interaction with NVs outside the sensing volume and reduced laser power requirements, benefiting scalability and efficiency~\cite{Yu2024,Solyom2024}. 
Collectively, these properties enable compact sensing architectures while also relaxing constraints on alignment and homogeneity of microwave and optical fields~\cite{Sorensen2025}.

\begin{figure}[b!]
    \centering
    \includegraphics[]{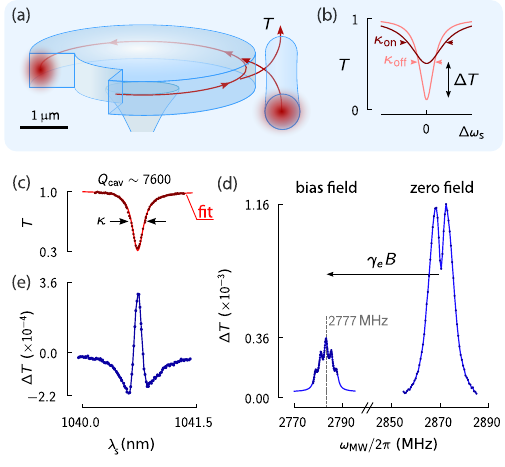}
    \caption{\textbf{Singlet-absorption ODMR in a nanocavity.} 
    {(a)} Coupling between a waveguide and a WGM cavity enables detection of spin-dependent changes in absorption by an NV ensemble, which manifest as changes in transmission, $\Delta T$ {(b)}. 
    $\kappa_{\text{on}}$ ($\kappa_{\text{off}}$) is the cavity linewidth with (without) a resonant microwave drive field and $\Delta\omega_{\text{s}} = \omega_{\text{s}}-\omega_{\text{cav}}$ is the IR probe frequency detuning.
    $\Delta T$ is exaggerated for clarity.
    {(c)} Coherent mode spectroscopy of the cavity reveals a WGM near the NV singlet transition. 
    The red line is a fit [\Cref{eq:Tcav}] to the transmission spectrum.
    {(d)} Sitting on resonance with the WGM, we perform ODMR with and without an axial magnetic bias field $B$.
    Hyperfine coupling to the $^{14}$N nuclear spin produces three transitions separated by $\Omega_{\text{HF}} = 2.16$~MHz~\cite{Smeltzer2009}.
    The five peaks produced in the presence of a bias field result from simultaneously driving all three hyperfine resonances of one of the four crystallographic orientations~\cite{Chakravarty2026},
    whereas the zero-field features arise from electric fields from nearby charged dopants and strain interactions~\cite{Sorensen2025}.
    The light blue line in (d) is a fit using \Cref{eq:fit_ODMR_3}.
    {(e)} We drive the spin transition at 2777\,MHz (d) while stepping the laser from $\lambda_{\text{s}} = 1040-1041.5$\,nm and measure $\Delta T(\lambda_{\text{s}})$. A line connecting the scatter points has been added for clarity.
    }
    \label{fig:fig2_ODMR}
\end{figure}

In this work, we realize singlet-absorption magnetometry in monolithic wavelength-scale diamond microdisk cavities [\Cref{fig:fig2_ODMR}~(a)]. The resulting chip-based sensors miniaturize previous demonstrations of ODMR based on single-pass~\cite{Acosta2010,Schall2025,TayefehYounesi2025} and cavity-enhanced~\cite{Dumeige2013,Jensen2014,Gottesman2024} singlet-absorption in macroscopic diamond samples.
This platform benefits from cavity-enhanced sensitivity, miniaturization and robustness; however, we also discover and elucidate additional physics that must be considered when implementing nanoscale diamond quantum magnetometers.
Specifically, \Cref{eq:sensitivity} suggests that increasing the input power $P_{\text{s}}$, thereby increasing $P_{\text{det}}$, can improve sensitivity without bound. However, high IR intensities, such as those achieved in nanoscale cavities for even modest input powers, can induce ionization from $^3\!E$ [\Cref{fig:fig1}~(b)]~\cite{Shandilya2025,Flagan2025,Itoi2026}.
Optimizing sensitivity therefore requires careful engineering of the optical environment to simultaneously maximize $C$ and $P_{\text{det}}$ while minimizing charge state conversion (CSC).
To achieve this, we develop a master-equation model that accurately predicts the spin-dependent cavity response and describes the optimal operating conditions and corresponding sensitivity, providing a roadmap for the future optimization of nanoscale singlet-absorption magnetometers.

\section{Singlet-absorption ODMR in a nanocavity}
\label{sec:sec2}
In singlet-absorption ODMR, spin-dependent absorption is measured by the transmission through a diamond sample of a probe laser resonant with the singlet transition.
In optical cavities, such as the diamond microdisk studied here [\Cref{fig:fig2_ODMR}~(a)], absorption modifies the cavity loss rate and, consequently, both the cavity resonance linewidth and contrast~\cite{Coutts2026}.
In nanophotonic cavities, these parameters can be monitored using a laser transmitted through a waveguide coupled to the cavity; we perform this cavity mode spectroscopy using a fibre-taper waveguide that evanescently transfers light into and out of the microdisk.

Loss in an optical cavity mode is characterized by its total energy decay rate $\kappa$, to which several sources of loss contribute~\cite{Coutts2026}:
\begin{align}
\kappa = \kappa_{\text{i}} + \kappa_{\text{p}} + \kappa_{\text{ex}} + \kappa_{\text{abs}}.
\label{eq:kappaComponents}
\end{align}
Here, $\kappa_{\text{i}}$ accounts for surface scattering, radiation loss, and non-NV absorption that is intrinsic to the device, $\kappa_{\text{ex}}$ for external coupling to the waveguide, $\kappa_{\text{p}}$ for parasitic coupling induced by the waveguide to undesired channels, and $\kappa_{\text{abs}}$ for absorption by NV centres. 
Spin-dependent changes in absorption modify $\kappa_{\text{abs}}$, which in turn modifies the cavity linewidth $\kappa$ and the resulting transmission $T(\Delta\omega_{\text{s}})$ [\Cref{fig:fig2_ODMR}~(b)], where $\Delta\omega_{\text{s}}$ is the detuning between the input laser and the cavity resonance frequency.
The spin-dependent change in transmission is defined as the transmission contrast $\Delta T$. 
There are two limiting regimes that describe the relative contribution of absorption loss.
In NV singlet-absorption-limited cavities ($\kappa_{\text{abs}}\gg\kappa_{\text{i}} + \kappa_{\text{p}}$), spin-dependent absorption strongly modifies the cavity linewidth and resonance depth.
Conversely, in scattering-limited cavities ($\kappa_{\text{abs}} \ll \kappa_{\text{i}} + \kappa_{\text{p}}$), spin-related absorption is weak, producing small ODMR signals. 
The device studied here operates in the latter regime.

In this work, we study a monolithic diamond nanocavity [\Cref{fig:fig2_ODMR}~(a)] with a thickness of $\sim800$~nm and a diameter of $\sim4.2\,\upmu$m. 
The cavity is one of several fabricated on the same chip, which support whispering-gallery modes (WGMs) with mode volumes below $2\,\upmu\text{m}^3$ and quality factors $Q>7\times10^4$~\cite{Coutts2026}.
The devices are fabricated from a dense-NV single-crystal diamond substrate (Element Six B14~\cite{ElementSix2021}) engineered to host NVs with a concentration of $4.5$ ppm while maintaining high spin-coherence times~\cite{Sorensen2025}.
Optical excitation and readout are performed entirely through a dimpled fibre-taper positioned adjacent to the cavity [\Cref{fig:fig2_ODMR}~(a)]~\cite{Mitchell2019,Masuda2024,Sorensen2025}, enabling compact waveguide-coupled operation without free-space optics (see Appendix~\ref{app:SI-setup}).
The same cavities have previously been used to investigate saturable absorption~\cite{Coutts2026} and triplet-PL magnetometry~\cite{Sorensen2025}, granting a direct comparison between readout modalities.

To probe spin-dependent absorption, we perform optical spectroscopy of the cavity while manipulating the spin state of the NVs embedded within it. 
IR laser light is evanescently coupled into the cavity from the fibre-taper, and we record the resulting transmission spectrum. 
Chosen for its spectral proximity to the singlet ZPL, we identify an under-coupled, scattering-limited WGM with quality factor $Q_{\text{cav}} = \omega_{\text{cav}}/(\kappa_{\text{i}} + \kappa_{\text{p}} + \kappa_{\text{abs}}) =  7.6\,(2)\times 10^3$ [\Cref{fig:fig2_ODMR}~(c)] and resonant frequency $\omega_{\text{cav}}/2\pi = 288.04~\text{THz}$ ($\lambda_{\text{cav}}=1040.8$~nm).
For spin readout, the NV ensemble is optically pumped through the fibre-taper with a cw 532~nm laser while the cw IR probe laser is held resonant with the WGM. 
The spin state is redistributed using an amplitude-modulated microwave field.
The modulated spin state and resulting changes to the singlet population modulate the level of cavity absorption.
The resulting spin-dependent changes in cavity transmission are demodulated using a lock-in amplifier referenced to the microwave modulation frequency. 
For details on the experimental setup and measurement techniques, see Appendices~\ref{app:SI-setup} and \ref{app:ODMRLITechniques}, respectively.
After experimentally optimizing the optical and microwave drive conditions (Appendix \ref{app:optimization}), we measure ODMR spectra as a function of microwave frequency, $\omega_{\text{MW}}$ [\Cref{fig:fig2_ODMR}~(d)].

\begin{figure*}[b!]
    \centering
    \includegraphics[]{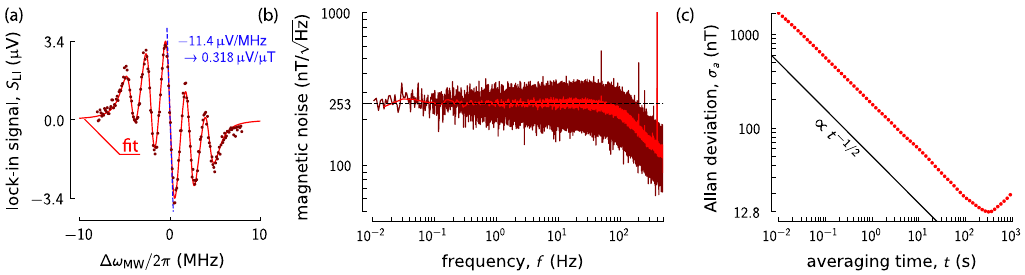}
    \caption{\textbf{Magnetic sensing and sensor stability.}
    {(a)} The measured ODMR spectrum reveals the point of highest sensitivity, whose slope is highlighted by the dashed blue line [the fit is produced using \Cref{eq:fit_ODMR_3_LI}]. 
    Driving the optimized spin ensemble at this point, we monitor $S_{\text{LI}}(t)$ over 1000~s and convert it into a magnetic field signal. 
    {(b)} The power spectral density of the magnetic time trace gives the magnetic noise (dark red line). The light red line is a smoothed signal. 
    The mean of the magnetic noise in the flat region gives the sensitivity of the magnetometer [$253$\,(9)\,nT$\,\text{Hz}^{-1/2}$]. 
    The uncertainty is determined from the fit shown in (a).
    {(c)} We use the Allan deviation of the time signal to determine the stability of the sensor, and, as demonstrated by the $t^{-1/2}$ dependence, the sensor is dominated by white noise at short time scales.
    }
    \label{fig:fig6_measSens}
\end{figure*}
As expected for a scattering-limited cavity, the maximum spin-dependent transmission contrast is modest, reaching $\Delta T = 3.6\times 10^{-4}$ for a single NV orientation [\Cref{fig:fig2_ODMR}~(d), left], approximately one quarter of the zero-field ODMR contrast obtained when all four crystallographic orientations contribute [\Cref{fig:fig2_ODMR}~(d), right]. 
By adding a magnetic bias field, we can Zeeman-split the spin transition frequencies, allowing us to address the spin state of a single NV orientation. 
From a fit to the ODMR spectrum with a magnetic bias field present, we extract a spin linewidth of $\nu = 1.82\,(2)$~MHz, comparable to that measured using triplet-PL ODMR in the same sample ($1.65\,(3)$~MHz)~\cite{Sorensen2025}.
Note that although other absorbers have been reported in the spectral range measured here~\cite{Fuchs1995,TayefehYounesi2025,Talik2026,Coutts2026}, we attribute the observed absorption primarily to the NV singlet transition due to its clear $\omega_{\text{MW}}$-dependent response.

Next, we explore the wavelength-detuning dependence of the absorption.
With a bias field applied, we fix $\omega_{\text{MW}}$ to the 2777~MHz transition and step the laser across the cavity resonance while measuring $\Delta T$.
As shown in \Cref{fig:fig2_ODMR}~(e), $\Delta T$ is maximized on resonance and decreases with detuning, eventually changing sign---all observations are consistent with an under-coupled cavity. 
In addition to spin-dependent absorption, a small contribution to $\Delta T$ arises from microwave-induced thermo-optic modulation of the cavity resonance frequency.
This dispersive contribution is much smaller than the spin-dependent absorption signal, as evidenced by the strong dependence of $\Delta T$ on $\omega_{\text{MW}}$ [\Cref{fig:fig2_ODMR}~(d)], although it could be responsible for some of the asymmetries seen in \Cref{fig:fig2_ODMR}~(e).
Regardless, these results demonstrate a contrast mechanism with which we can perform magnetic field sensing.

\section{Experimental magnetic sensitivity of the nanocavity}
\label{sec:sec3}

Here, we measure the magnetic sensitivity of the nanocavity. 
Although pulsed singlet-absorption protocols could ultimately provide higher sensitivity, as in triplet-PL ODMR~\cite{Fang2013,Barry2020,Sorensen2025}, we employ cw ODMR because of its experimental simplicity and compatibility with low-noise lock-in detection. 
Continuous-wave green illumination also suppresses IR absorption from other point defects~\cite{Collins2005,Khan2013,Ashfold2020,Johnson2025,Talik2026}, which was shown to be a significant source of loss in the diamond material studied here~\cite{Coutts2026}.

After optimizing the optical and microwave parameters, we implement a frequency-modulated microwave drive, which yields improved stability over amplitude modulation by suppressing microwave-induced thermo-optic fluctuations and by requiring a smaller dynamic lock-in range (Appendix~\ref{app:ODMRLITechniques}).
The resulting lock-in ODMR spectrum is shown in \Cref{fig:fig6_measSens}~(a).
From a fit to the ODMR spectrum (Appendix \ref{app:fits}), we extract the point of maximum slope and the corresponding calibration factor $\partial S_{\text{LI}}/\partial B = 0.318\,(11)\,\upmu\text{V}/\upmu\text{T}$ that relates the change in magnetic field to a change in lock-in signal $S_{\text{LI}}$.
Fixing the microwave detuning at this operating point, we record the time trace $S_{\text{LI}}(t)$ over 1000\,s and convert it to a magnetic field signal via the calibration factor. 
The power spectral density of this time series yields a direct current (dc) magnetic sensitivity of $253\,(9)\,\text{nT}\,\text{Hz}^{-1/2}$ ($\sim300\,\text{nT}\,\upmu\text{m}^{3/2}\,\text{Hz}^{-1/2}$), as shown in Fig.~\ref{fig:fig6_measSens}~(b). 

The long-term stability is characterized using the Allan deviation of the magnetic time trace [\Cref{fig:fig6_measSens}~(c)].
The Allan deviation exhibits an inflection near $t\sim30$~s, marking the onset of drift-limited performance that we attribute primarily to slow fluctuations in laser power and cavity temperature~\cite{Huang2019-2,Panuski2020}.
Its minimum sets the drift-limited minimum measurable field of $12.8\,(4)$\,nT, which could be improved through active stabilization of the cavity and laser~\cite{Dacha2025}.

\begin{figure*}[b!]
    \centering
    \includegraphics[]{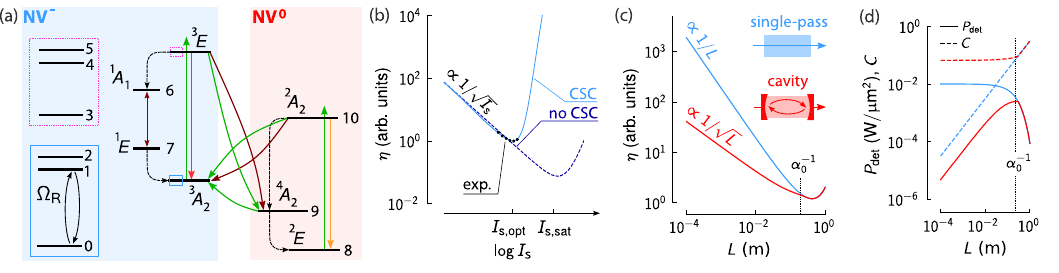}
    \caption{\textbf{Modelling the sensitivity of singlet-absorption ODMR.} 
    {(a)} Simplified eleven-state diagram of the NV centre. 
    Solid (dashed) lines indicate the primary optical (non-radiative) transitions. 
    Charge-state conversion between NV$^{-}$ and NV$^0$ can be caused by the intensity-dependent absorption of both green and IR light; therefore, $p_7$, which primarily determines the spin-dependent NV absorption coefficient, is also intensity-dependent.
    {(b)} At low $I_{\text{s}}$, $\eta$ improves as $I_{\text{s}}^{1/2}$; at high intensity, however, power broadening and CSC degrade $\eta$ before saturation of the singlet occurs ($I_{\text{s}} = I_{\text{s,sat}}$). 
    The achievable sensitivity is overestimated (dashed lines) if CSC is not considered.
    The measured sensitivity is shown for comparison (both normalized). 
    {(c)} The sensitivity of a small single-pass sensor improves as $1/L$. 
    The sensitivity of an etalon outperforms the single-pass sensor by a factor of $(\alpha_0 \, L)^{-1/2}$ in the small-sensor limit. 
    The optical penetration depth $(\alpha_0^{-1})$ is highlighted by the dotted line.
    (d) A comparison of the $C$ and $P_{\text{det}}$ for a single-pass sensor and an ideal cavity sensor reveals that a cavity improves $C$ at the expense of $P_{\text{det}}$.
    }
    \label{fig:fig4_model}
\end{figure*}
Equation~\ref{eq:sensitivity} predicts a PSNL sensitivity of $\eta_{\text{PSNL}} = 81\,(12)\,\text{nT}\,\text{Hz}^{-1/2}$ (see Appendices~\ref{app:highQ} and \ref{app:comparison} for calculation details), a factor of $\sim3$ better than the measured value. 
We primarily attribute this discrepancy to intensity noise of both the pump and probe lasers and fast temperature fluctuations of the cavity resonance.
Although we stabilize the cavity using a thermoelectric cooler, residual temperature fluctuations shift the cavity resonance relative to the probe laser via thermo-optic effects, modulating the transmission independently of the microwave field. 
These effects are partially mitigated by the large cavity linewidth---higher $Q$ cavities are more susceptible to noise caused by thermal fluctuations and would require more extensive stabilization techniques to counter the additional noise sources~\cite{Sansa2016,Roy2018}. 

To test whether these sources of noise limit the sensitivity of higher $Q$ cavities, we characterize a second microdisk with a slightly larger diameter ($4.4\,\upmu$m) supporting a higher $Q$ mode ($Q_{\text{cav}}\sim35\times10^3$) near the singlet ZPL ($1041.4$~nm). 
Although the device produced both larger $\Delta T$ and a similar PSNL sensitivity estimation [$\eta_{\text{PSNL}}=89\,(14)\,\text{nT}\,\text{Hz}^{-1/2}$] to the other cavity WGM due to worse fibre-taper coupling, the measured sensitivity was only $\eta=368\,(12)\,$nT$\,{\text{Hz}}^{-1/2}$.
This behaviour is consistent with increased susceptibility of higher $Q$ cavities to temperature fluctuations~\cite{Sansa2016,Roy2018}, which outweigh the improvement in ODMR contrast. 
See Appendix~\ref{app:highQ} for details on these higher $Q$ measurements. 

Lastly, we also compare the magnetic sensitivity achievable using cavity-collected singlet PL, as shown by the measurements in Appendix~\ref{app:singletPLODMR}. 
Despite exhibiting a zero-field ODMR contrast of approximately $20\,\%$ [\Cref{fig:fig8_SingletPL}~(a)], singlet-PL readout is fundamentally limited by the low radiative quantum efficiency of the singlet transition, resulting in an estimated PSNL sensitivity of $\sim10\,\upmu\text{T}\,\text{Hz}^{-1/2}$ that is more than an order of magnitude worse than singlet-absorption ODMR. 

\section{A model to predict the optimal sensitivity}
\label{sec:section4}
A primary goal of this work is to determine the achievable magnetic sensitivity using singlet-absorption ODMR in nanophotonic cavities.
Previous theoretical studies~\cite{Dumeige2013,Zadeh-Haghighi2024} have predicted sub-pT$\,\text{Hz}^{-1/2}$ sensitivities at sub-millimetre scales; however, these models consider only spin-conserving transitions or neglect CSC induced by intense IR fields, which can be reached inside nanocavities using modest input powers~\cite{Coutts2026} but have only recently been studied in this context \cite{Shandilya2025}.
Here we address these limitations and test the resulting model against our experimentally demonstrated nanophotonic cavity sensor.

In Section~\ref{sec:section4.1}, we describe a master-equation model of the NV centre that enables the prediction of ODMR linewidth and the optimal operating conditions for magnetic sensing. 
In Section~\ref{sec:section4.2}, we detail how singlet absorption is transduced into a singlet-absorption ODMR signal, enabling the prediction of the magnetic sensitivity for both single-pass and cavity geometries.

\subsection{NV population dynamics}
\label{sec:section4.1}
The contrast mechanism in singlet-absorption ODMR arises from the spin-dependent absorption coefficient of the NV ensemble, $\alpha$~\cite{Zadeh-Haghighi2024}: 
\begin{align}
    \alpha =\xi \,M\, p_{7}\,\sigma^{\text{s}}_{7,6}\,,
    \label{eq:absCoefficient}
\end{align}
where $p_7$ is the steady-state spin-dependent population density of the metastable state $^1\!E$ [\Cref{fig:fig4_model}~(a)], $\sigma^{\text{s}}_{7,6}$ is the singlet absorption cross-section, $M$ is the NV density, and $\xi$ is a scaling factor used to correct for polarization mismatch between the optical field and the NV ensemble (see Appendix~\ref{app:MasterEquationModel}).
The cw ODMR contrast is determined by the change in steady-state absorption between the microwave-on and microwave-off states, $\Delta\alpha$, which is set by the corresponding change in $p_7$, $\Delta p_7$. 
Consequently, the achievable contrast depends not only on the microwave drive, which induces Rabi flopping ($|0\rangle\rightarrow|1,2\rangle$) at frequency $\Omega_{\text{R}}$, but also the intensities of the green and IR optical fields that govern the steady-state population dynamics, respectively $I_{\text{p}}$ and $I_{\text{s}}$.

Previous models of singlet-absorption magnetometry either considered only spin-conserving optical transitions or did not consider the effects of multi-photon absorption and CSC~\cite{Dumeige2013,Zadeh-Haghighi2024,Wang2026}.
More recent experiments have demonstrated that intense IR fields can substantially modify the NV population dynamics and charge-state distribution, making these effects essential for accurately predicting sensor performance~\cite{Shandilya2025}. 
To that end, we extend the master-equation approach taken by Zadeh-Haghighi et al.~\cite{Zadeh-Haghighi2024} (see also Ref. \cite{Wirtitsch2023}).

To predict the ODMR response under a wide range of operating conditions, we solve the Lindblad master equation at steady state for the model of a single NV [see \Cref{fig:fig4_model}~(a) for a simplified illustration] with $T_2^*\approx0.43\,\upmu$s across a range of experimental parameters ($\Omega_{\text{R}},\,I_{\text{p}},\,I_{\text{s}}$); we choose $T_2^*$ to match that of the NV ensemble in our cavity~\cite{Sorensen2025}.
The intensity-dependent transition rates result in steady-state populations that vary greatly with $\Omega_{\text{R}},\,I_{\text{p}}$, and $\,I_{\text{s}}$.
The optimal set ($\Omega_{\text{R,opt}},\,I_{\text{p,opt}},\,I_{\text{s,opt}}$) maximizes the expression $\sqrt{I_{\text{s}}}\Delta p_7/\nu$, where $\nu$ is the ODMR linewidth, enabling us to predict the optical and microwave operating conditions that maximize the magnetic sensitivity for the dense-NV material studied here. 
The optimal parameters are tabulated in \Cref{tab:tab7_OptParametersMain} alongside $\nu$, $\Delta\alpha$, and the nominal absorption $\alpha_0$ predicted at those conditions.
Full details of the model can be found in Appendix~\ref{app:MasterEquationModel}.

\setlength{\extrarowheight}{2pt}
\begin{table}[t!]
\centering
\caption{Summary of optimized parameters predicted by the model and experimentally measured for an NV ensemble with $T_2^*\approx0.43\,\upmu$s. 
The model-predicted PSNL sensitivity accounts for detector and fibre transmission loss and assumes an effective mode volume of $V_{\text{eff}} \approx 1.6\,\upmu\text{m}^3$\,\cite{Coutts2026}. Further, we assume that the WGM's group index matches the refractive index of diamond.
Note that we cannot reliably determine the green intensity inside the cavity.
$\dagger$~The nominal experimental NV contribution to absorption cannot be separated from other absorptive loss. }
\begin{tabular*}{\linewidth}{@{\extracolsep{\fill}}cccc}
parameter & model & experiment & unit \\ \hline \hline 
$\Omega_{\text{R,opt}}/2\pi$ & 316   & $\sim 500$ &kHz \\
$I_{\text{p,opt}}$  & 0.0314& ---- &MW/cm$^2$ \\
$I_{\text{s,opt}}$ & 1.01& $1.3\,(4)$ &MW/cm$^2$  \\
$\nu$ & 1.81 & $1.82\,(2)$ & MHz \\
$\Delta\alpha$ & 0.36& $0.3\,(1)$& m$^{-1}$ \\
$\alpha_0$$^\dagger$ & 4.77& ---- & m$^{-1}$ \\
$\eta_{\text{PSNL}}$$^\ddagger$ & 66& $81\,(12)$& nT$\,\text{Hz}^{-1/2}$ \\
\hline \hline
\label{tab:tab7_OptParametersMain}
\end{tabular*}
\begin{minipage}{\linewidth}
\footnotesize
$^\ddagger$~Estimated using the experimentally determined $I_{\text{s}}$, $\nu$, and $C$. 
\end{minipage}
\end{table}

The principal improvement of our model over previous work lies in the consideration of single- and multi-photon absorption processes outside the singlet-manifold transition. 
To illustrate the importance of this consideration, we calculate the normalized sensitivity of a singlet-absorption magnetometer as a function of $I_{\text{s}}$. 
As shown in \Cref{fig:fig4_model}~(b), the sensitivity reaches an optimum at a finite intensity $I_{\text{s,opt}}$ and degrades at higher $I_{\text{s}}$. 
If only singlet-transition saturation is considered (no CSC), the achievable sensitivity is substantially overestimated.
This result is particularly relevant for nanophotonic sensors, which can routinely achieve high intensities with low input powers~\cite{Coutts2026}, as demonstrated by the intensity dependence of the experimentally measured cavity sensitivity [\Cref{fig:fig4_model}\,(b)].
CSC fundamentally limits the usable probe intensity~\cite{Shandilya2025} and, therefore, the achievable magnetic sensitivity. 

The relevant optical transitions, their rates, and the optimal conditions are sensitive to the frequency of the fields present; however, our model could in principle be extended to solve for the optimal conditions given other pump or probe wavelengths, pulsed modalities~\cite{Hopper2016}, or those based on other detection mechanisms like spin-to-charge conversion readout~\cite{Shields2015}.
These would require further studies on NV charge-state dynamics in the presence of other optical frequencies.
The same methods can be used to find the optimal conditions for different diamond materials (i.e. different $M$ and $T_2^*$).

\subsection{Cavity-enhanced readout}
\label{sec:section4.2}
The preceding model of the NV population dynamics determines the optical and microwave conditions that maximize the spin-dependent absorption response of the NV ensemble.
The magnetic sensitivity, however, additionally depends on how the absorption is transduced into a measurable optical signal, and is therefore geometry-dependent.
To establish the role of cavity enhancement in this transduction, we first consider a single-pass transmissive sensor, where the only source of loss is absorption by NV centres. 
For a small sensor, whose length $L$ is much smaller than the penetration depth (for the material studied here, $\,L\ll \alpha_0^{-1}\approx21\,$cm) and whose probe intensity matches the optimum value determined above ($I_{\text{s,opt}}$), the PSNL sensitivity is given by (Appendix~\ref{app:comparison})
\begin{align}
    \eta_{\text{single}} \approx \beta\frac{\nu}{\sqrt{I_{\text{s,opt}}\,A}}\frac{1}{\Delta\alpha\,L},
    \label{eq:singlePassMainSens}
\end{align}
as shown in \Cref{fig:fig4_model}\,(c).
Here, $A$ is the cross-sectional area of the probe laser inside the diamond.
As shown in \Cref{fig:fig4_model}~(d), increasing the length of small, single-pass sensors increases the number of participating NV centres and the optical contrast, correspondingly improving the sensitivity.
This improvement saturates as $L\rightarrow\alpha_0^{-1}$, beyond which the reduction in $P_{\text{det}}$ from exponential absorption offsets the increase in $C$. 
This exponential absorption is responsible for the inflection point in $\eta$ in \Cref{fig:fig4_model}~(c).

Embedding a small sensor in an optical cavity can enhance the effective interaction between the probe field and the NV ensemble, compensating for the reduced length~\cite{Dumeige2013}. 
This enhanced interaction increases $C$ while decreasing $P_{\text{det}}$, since the cavity-enhanced $I_{\text{s,opt}}$ is reached for a lower input power [\Cref{fig:fig4_model}~(d)]. 
Overall, however, the exchange between $C$ and $P_{\text{det}}$ improves the sensitivity.
In the small-sensor limit, both optimized symmetric Fabry--Perot etalons and side-coupled cavities like the one measured here, improve upon a single-pass sensor of the same length by
\begin{align}
\frac{\eta_{\text{single}}}{\eta_{\text{cav}}}
\approx \frac{1}{\sqrt{\alpha_0L}},
\label{eq:improvement}
\end{align}
assuming that the only cavity loss is optical absorption by NVs, that the intracavity intensity equals $I_{\text{s,opt}}$, and that the cavity is nearly critically coupled (see Appendix~\ref{app:comparison}).
Note, however, that while a cavity can improve the sensitivity of a short sensor,
a nanocavity containing a small NV ensemble cannot ultimately outperform a sufficiently long single-pass sensor containing a large NV ensemble [\Cref{fig:fig4_model}~(c)].

\begin{table*}[b!]
\centering
\caption{Projected improvements to the sensitivity. Starting from the measured sensitivity, each row incorporates an additional method of improvement.
The columns give the projected sensitivity, the improvement factor from that step alone, and the cumulative improvement relative to experiment. 
}
    \begin{tabular}{cccc}
& sensitivity (nT\,$\text{Hz}^{-1/2}$) & individual improvement& cumulative improvement \\ \hline \hline 
experiment & 253 & ---&---   \\ \hline
stabilization (PSNL) & $81$  & $3.1$&$3.1$   \\ \hline
unity probe collection & $31$  & $2.6$&$8.2$   \\ \hline
NV-absorption-limited $Q$ & $1.1$& $28$ & $230$    \\ \hline
pulsed protocols$^\dagger$ & $\sim0.2$ & $\sim5$ & $\sim10^3$ \\ \hline
isotopic purification~\cite{Ishikawa2012} & $\sim0.1$ & $\sim2$ & $\sim2\times10^3$ \\ \hline
spin-bath driving \& double-quantum~\cite{Bauch2018} & $\sim0.01$ & $\sim10$ & $\sim2\times10^4$ \\ \hline \hline
\label{tab:improvement}
\end{tabular}
\begin{minipage}{\linewidth}
\raggedright
\small
$^\dagger$\,Modelled improvement to the ODMR contrast and probe intensity.
\end{minipage}
\end{table*}

Furthermore,  as described in Eq.\ \eqref{eq:kappaComponents}, real cavities exhibit sources of loss unrelated to absorption by NV centres that can be larger than NV absorption loss ($\kappa_{\text{i}} +\kappa_{\text{p}} \gg \kappa_{\text{abs}}$), thereby degrading the ODMR contrast. 
In such a scattering-limited, small cavity sensor, $\kappa\gg \Delta\kappa$, and the sensitivity can be expressed as (Appendix~\ref{app:comparison})
\begin{align}
    \eta_{\text{cav}} \propto \frac{\kappa}{\Delta\kappa\sqrt{L}}.
    \label{eq:etaCavProp}
\end{align}
This scaling highlights the need to mitigate sources of non-singlet-absorption loss, thereby reducing $\kappa$. 
Equation $\ref{eq:etaCavProp}$ also highlights the importance of NV density ($\Delta\kappa\propto M$) and cavity length, each of which impacts the amount of absorption.

\begin{figure}[t!]
    \centering
    \includegraphics[]{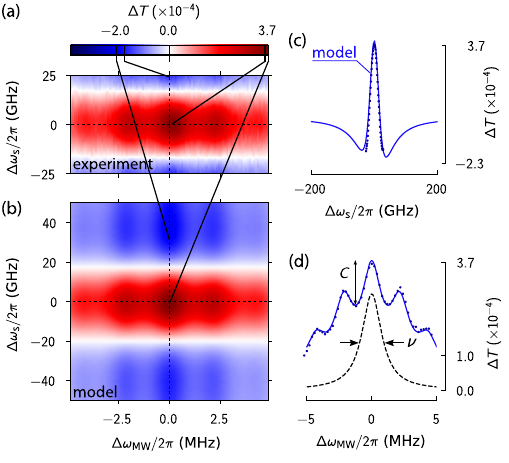}
    \caption{\textbf{Modelling the transmission contrast. }
    Over the full two-dimensional parameter space, measurements (a) of $\Delta T(\Delta\omega_{\text{MW}},\Delta\omega_{\text{s}})$ show excellent agreement with the model predictions (b). 
    We use the parameters extracted from a fit to the transmission spectrum [\Cref{fig:fig2_ODMR}(c)] and the model to calculate $\Delta T(\Delta\omega_{\text{MW}},\Delta\omega_{\text{s}})$. 
    (c) Vertical line cuts of (a) and (b) compare the measured and predicted $\Delta T(\Delta\omega_{\text{s}})$; (d) horizontal line cuts of (a) and (b) compare the measured and predicted ODMR spectra. 
    }
    \label{fig:fig7_modelComp}
\end{figure}

The full cavity model (Appendix~\ref{app:comparison}) combines the NV absorption model from Section~\ref{sec:section4.1} with the cavity-waveguide coupled-mode theory to predict the spin-dependent $\Delta T$ of a non-ideal cavity.
The intensity-dependent NV absorption coefficients $\alpha^{(\text{on})}$ and $\alpha^{(\text{off})}$ are calculated from the master-equation model using independently determined quantities: the NV density $M$ and inhomogeneous dephasing time $T_2^*$, the intensity-dependent transition rates and absorption cross-sections measured in previous studies, and the experimental Rabi frequencies and optical intensities. 
Moreover, the cavity parameters are obtained from a fit to the fibre-taper transmission spectrum of the studied WGM [\Cref{fig:fig2_ODMR}~(c)]. 
No other parameter is fitted. 
With these inputs, we use the model to predict $\Delta T$ as a function of $\Delta\omega_{\text{s}}$ and the microwave detuning from spin resonance $\Delta\omega_{\text{MW}}$.
Over the full two-dimensional space $\Delta T(\Delta\omega_{\text{MW}},\Delta\omega_{\text{s}})$, the measured  [\Cref{fig:fig7_modelComp}~(a)] and predicted [\Cref{fig:fig7_modelComp}~(b)] maps are in excellent agreement, and the experimentally optimized $I_{\text{s}}$, $\Delta\alpha$, and $\nu$ each agree within uncertainty of the model predictions (\Cref{tab:tab7_OptParametersMain}).
The agreement is highlighted by the line cuts shown in  \Cref{fig:fig7_modelComp}~(c) and \Cref{fig:fig7_modelComp}~(d).
The optical detuning dependence is a particularly good test of the model, given that $I_{\text{s}}$, and therefore $\alpha$, change as the probe is detuned from resonance, modifying the $\Delta T$ lineshape. 
We see similarly good agreement when performing magnetometry with a second higher-$Q$ doublet mode in a different cavity, again with no free parameters (Appendix~\ref{app:highQ}).
The excellent agreement across the full two-dimensional parameter space demonstrates that the model accurately captures both the cavity response and the intensity-dependent absorption dynamics.

\section{The ultimate sensitivity limits of NV ODMR}
\label{sec:section5}

\begin{figure}[b!]
    \centering
    \includegraphics[]{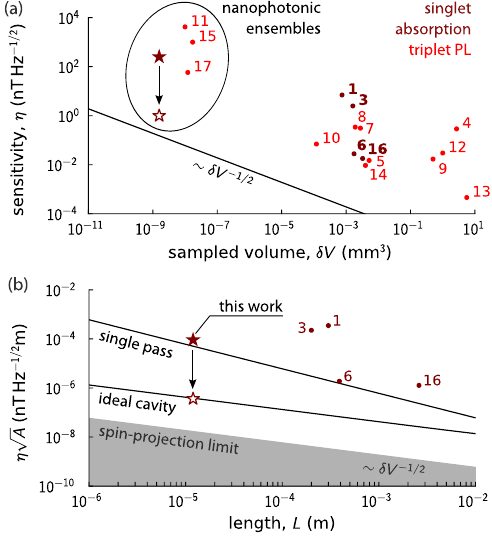}
    \caption{\textbf{Comparison of experiment and model-predicted limits.} 
    {(a)} Comparison of the sensitivity as a function of sensing volume for different experimental magnetometers: 1\,\cite{Acosta2010}; 2\,\cite{Fang2013}; 3\,\cite{Jensen2014}; 4\,\cite{Clevenson2015}; 5\,\cite{Barry2016}; 6\,\cite{Chatzidrosos2017}; 
    7\,\cite{Patel2020}; 8\,\cite{Strner2021}; 9\,\cite{Zhang2021}; 10\,\cite{Shim2022}; 11\,\cite{Li2023}; 12\,\cite{Graham2023}; 13\,\cite{Barry2024}; 14\,\cite{Sekiguchi2024}; 15\,\cite{Katsumi2025}; 16\,\cite{TayefehYounesi2025}; 17~\cite{Sorensen2025}.  
    The stars denote the results of this work. 
    Light and dark red numbers highlight triplet-PL and singlet-absorption magnetometers, respectively.
    The singlet-absorption sensors are bolded for clarity. 
    The black arrow highlights the sensitivity improvement assuming our cavity can be made absorption-limited.
    Sensitivity limits generally scale with $\delta V^{-1/2}$, where $\delta V$ is the sensing volume or, equivalently, the volume of diamond that contributes to the absorption.
    In our experiment, we define $\delta V$ as the optical mode volume.
    {(b)} We compare experimental absorption-based magnetometers to the area-normalized sensitivity limits predicted by our model both with and without an ideal cavity (Appendix~\ref{app:comparison}). 
    The single-pass and ideal cavity lines are calculated respectively using \Cref{eq:etacwBulk,eq:etaFPcav}.
    The material-dependent spin-projection limit is calculated as $\eta_{\text{sp}}= (\gamma_e \sqrt{NT_2^*})^{-1}$, where $N$ is the sample volume multiplied by the NV density and $T_2^*=0.43\,\upmu$s is the inhomogeneous dephasing time~\cite{Sorensen2025}. 
    }
    \label{fig:fig6_litReview}
\end{figure}

The ultimate sensitivity of an NV magnetometer is limited by spin-projection noise~\cite{Budker2007}.
In practice, however, optical readout introduces additional noise because the spin state is inferred indirectly from the rate of optical detection~\cite{Barry2020}. 
The properties of the NV centre therefore set a sensitivity limit, and the gap between this limit and experiment quantifies the remaining scope for improvement.

Having validated the model experimentally, we use it to determine this practical sensitivity limit.
Assuming that a stabilized cavity with the same dimensions and material as the one studied here can be made NV-absorption-limited ($Q_{\text{cav}}\approx5\times10^5$)~\cite{Mitchell2019}, and that the waveguide and detection efficiencies can be made unity, \Cref{eq:singlePassMainSens,eq:improvement} predict that our platform could reach a PSNL sensitivity of 1.1~nT$\,\text{Hz}^{-1/2}$ ($\sim1\,\text{nT}\,\upmu\text{m}^{3/2}\,\text{Hz}^{-1/2}$ normalized to sensor volume). 
The necessary sensor improvements to achieve this limit, which is only $\sim20$ times higher than the spin-projection limit ($\eta_{\text{sp}}\approx48$~pT~Hz$^{-1/2}$), are summarized in \Cref{tab:improvement}.
A sensor at this limit would have the best volume-normalized sensitivity of any NV magnetometer to date~\cite{Rovny2024}.
The improvement is illustrated using the black arrow in \Cref{fig:fig6_litReview}~(a), which compares our experimental results and the predicted limit of our platform with representative experimental demonstrations of ensemble magnetometers based on both singlet absorption~\cite{Acosta2010,Jensen2014,Chatzidrosos2017,TayefehYounesi2025} and triplet PL~\cite{Fang2013,Clevenson2015,Barry2016,Patel2020,Strner2021,Zhang2021,Shim2022,Li2023,Graham2023,Barry2024,Sekiguchi2024,Katsumi2025,Sorensen2025}. 
The importance of cavity enhancement is illustrated in \Cref{fig:fig6_litReview}~(b), which demonstrates that a cavity enhances the sensitivity of short sensors much more than longer ones.

The data in \Cref{fig:fig6_litReview}~(a) provoke a comparison of the limits of each ODMR readout modality, which is enabled by our model.
Similar to singlet-absorption ODMR, we optimize the experimental conditions ($\Omega_{\text{R}}, I_{\text{p}}$) for a triplet-PL ODMR scheme.
Our model predicts that, under optimal conditions, cw singlet-absorption ODMR outperforms cw triplet-PL ODMR by a factor of two, assuming unity collection efficiency for both schemes.
This advantage therefore arises from the intrinsic properties of the readout mechanisms rather than differences in collection efficiency, which can vary widely depending on the experimental geometry~\cite{Babinec2010,Hadden2010,LeSage2012,Wolf2015,Zhang2018,Strner2021,Graham2023,Barry2024,Guo2025}. 
In fact, efficient collection is much more feasible for a singlet-absorption scheme.
In this work, we measure a post-cavity laser detection efficiency of $\sim15\,\%$, accounting for fibre-taper and transmission losses, whereas the detection efficiency of triplet PL from the same cavity was estimated to be $<1\,\%$~\cite{Sorensen2025}. 
Furthermore, we infer an experimental photon flux of $\sim2\times10^{12}\,\text{s}^{-1}$, nearly 3 orders of magnitude higher than triplet PL~\cite{Sorensen2025}, despite using the same fibre-taper collection geometry and detector.

These advantages enable the performance of singlet-absorption ODMR [$253\,(9)\,\text{nT}\,\text{Hz}^{-1/2}$] to match that of triplet-PL ODMR ($220\,(40)\,\text{nT}\,\text{Hz}^{-1/2}$~\cite{Sorensen2025}, obtained from a similar nanocavity on the same chip) despite several experimental limitations that only affect singlet absorption. 
These limitations include low cavity $Q$ and poor stability. 
Although cavities provide a route to substantially enhanced sensitivity, they introduce additional technical noise sources absent in single-pass and triplet-PL sensors, including resonance-frequency fluctuations and sensitivity to laser detuning~\cite{Huang2019-2,Panuski2020}. 
Active cavity locking should substantially reduce these effects~\cite{Dacha2025}.

Overall, there are three main requirements to enable absorption-based nanophotonic sensors to reach the predicted limits: (i) absorption-limited cavities made from material with higher spin density, (ii) cavity resonance stabilization, and (iii) improvements in waveguide coupling and loss. Furthermore, isotopic purification~\cite{Ishikawa2012}, pulsed protocols, and spin-bath driving~\cite{DeLange2012,Knowles2013,Bauch2018} could together yield another two orders of magnitude in sensitivity improvement (\Cref{tab:improvement}).
In particular, the implementation of pulsed protocols (\Cref{tab:improvement}) could enable longer coherence times~\cite{Barry2020}; pulsed IR schemes might avoid charge-state conversion effects observed in steady-state cw experiments~\cite{Shandilya2025}, allowing the use of higher intensities to achieve higher ODMR contrast. 
This is a subject of future work.

Although optical readout remains fundamentally limited by photon statistics, cavity-enhanced singlet absorption narrows the gap between the achievable experimental sensitivity and the spin-projection limit, providing a practical path toward quantum-limited nanophotonic magnetometry.
These sensors could enable a variety of future applications including biomagnetic sensing~\cite{Krber2016,Barry2016}, nuclear magnetic resonance spectroscopy~\cite{Glenn2018,Liu2022,Neuling2023,Briegel2025}, and two-dimensional magnetic imaging~\cite{Maletinsky2012,Thiel2019,Zhou2020,Welter2022,Scheidegger2022,Li2023}, which could benefit from the small size of the investigated platform. 
Such applications would require advances in packaging, enabling the replacement of the fibre-taper waveguide with integrated on-chip waveguides.
Other nanophotonic cavity platforms~\cite{Itoi2026,Zhu2026} might also be suitable for such applications, necessitating future investigation. 
For more discussion of these applications, see Ref.~\cite{Sorensen2025}.

\section{Conclusions}
In summary, we have demonstrated singlet-absorption ODMR and magnetic field sensing in dense-NV diamond nanophotonic cavities. 
We measure a dc sensitivity of $253\,(9)\,\text{nT}\,\text{Hz}^{-1/2}$ and a drift-limited minimum measurable field of 12.8~(4)~nT.
This constitutes the first demonstration of a nanophotonic singlet-absorption magnetometer. 
The cavity platform enables efficient delivery and collection of both pump and probe fields using a compact fibre-coupled architecture while enhancing the optical interaction between the probe field and the NV ensemble. 
Fibre delivery removes the need for free-space optics and enables low laser power operation.
Moreover, singlet-absorption magnetometry enables high readout efficiency---we collect roughly 1000 times more light than traditional triplet-PL-based measurements.

In parallel, we develop a master-equation model that incorporates spin dynamics, singlet absorption, and IR-induced charge-state conversion to predict the performance limits of cavity-enhanced singlet-absorption magnetometry. 
The model accurately reproduces the measured cavity transmission and ODMR response and reveals that optimizing the optical environment requires balancing optical and microwave power, cavity enhancement, and cavity stability. 
The model, which has no free parameters, is validated by characterizing sensor performance using two cavities, and predicts an optimal photon-shot-noise-limited sensitivity that agrees with measured values.
Using this framework, we show that absorption-limited cavities with improved coupling and stabilization could improve the sensitivity of the demonstrated platform by more than 2 orders of magnitude, enabling sub-nT$\,\text{Hz}^{-1/2}$ sensitivities, a factor of only $20$ above the spin-projection limit.

These results demonstrate that IR-induced charge-state conversion sets the optimal probe intensity and the practical limit for singlet-absorption magnetometry. 
This work establishes cavity-enhanced singlet absorption as a promising route toward compact, fibre-coupled, chip-scale quantum sensors and provides a framework for engineering future NV magnetometers that operate closer to the fundamental quantum limit.

\subsection*{Acknowledgements}
We gratefully acknowledge helpful discussions and resulting suggestions from Lilian Childress. 
Additionally, we thank the staff at the Nanofab at the University of Alberta for their time and expertise.
This work was supported by NSERC (Discovery Grant program), NSERC Alliance-Alberta Innovates Advance Program, and the Canadian Foundation for Innovation. S.F. acknowledges support from the Swiss National Science Foundation (Project No. P500PT\_206919).

\subsection*{Competing interests}
The authors declare no competing interests. 

\subsection*{Data availability}
There are no publicly available research data or software supporting this manuscript. Requests for further information or data should be sent to the authors.

\appendix

\section*{Appendix}
\section{Experimental setup}
\label{app:SI-setup}

\begin{figure}[t!]
    \centering
    \includegraphics{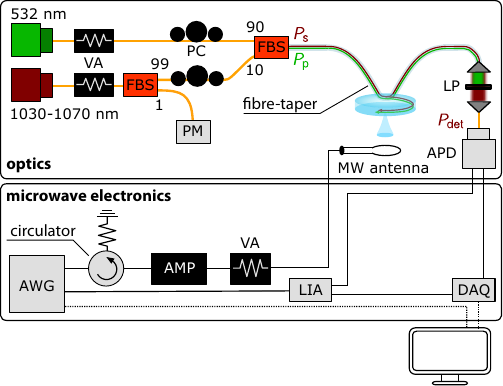}
    \caption{\textbf{Experimental setup.} 
    Two cw lasers---one green (532\,nm) and one IR (1030-1070\,nm)---are attenuated and injected into the microdisk using a dimpled fibre-taper. 
    The optical power of each laser is controlled using a variable attenuator. 
    The spin state of the NV ensemble is controlled using a microwave field generated using an arbitrary waveform generator and an antenna. 
    Spin-dependent IR cavity transmission is monitored using a photodiode after the green laser is filtered out using a long pass filter.
    List of abbreviations: AMP---amplifier; APD---avalanche photodiode; AWG---arbitrary waveform generator; DAQ---data acquisition tool; FBS---fibre beamsplitter; LIA---lock-in amplifier; LP---long pass filter; MW---microwave; PC---polarization controller; PM---power meter; SP---short pass filter; VA---variable attenuator. 
    }
    \label{fig:fig7_expSetup}
\end{figure}

The experimental setup used is shown schematically in \Cref{fig:fig7_expSetup}.
Control and readout of the NV ensemble are performed using two cw lasers---one visible (CrystaLaser LC CL532-025-SO, $532$\,nm) and one IR (New Focus Velocity TLB-6721, $1030-1070$\,nm)---that are evanescently coupled to the microdisk using a fibre-taper. 
The IR laser is injected into a 99:1 fibre beamsplitter (Thorlabs TW560R1A1), which directs 1~\% of the light to a power meter (Thorlabs PM400/S120C). 
The other 99\,\% passes to another fibre beamsplitter, which is used to combine the two laser fields.
The co-propagating laser fields pass into the fibre-taper~\cite{Michael2007,Masuda2024} (SMF-28) and evanescently scatter into the microdisk~\cite{Sorensen2025}. 
The intracavity IR field then couples back into the fibre-taper and is detected using an avalanche photodiode (Thorlabs APD410A/M).
Green light and NV PL are filtered from the IR probe light using a  633\,nm long-pass filter (Semrock BLP01-633R-25) and a 980 nm long-pass filter (Semrock BLP01-980R-25).

The chip is positioned optimally relative to the fibre-taper using two stepper-motor translation stages (Suruga Seiki XXC06020-G). 
The taper was then brought into contact with the side of the disk, choosing the position to optimally couple to the relevant WGM.
Once in contact, the taper position was fixed, ensuring constant fibre–cavity coupling throughout the experiment.
To avoid undesired contact between the fibre-taper and the diamond substrate, the fibre-taper was fabricated with a dimple\,\cite{Michael2007,Masuda2024}.
The results described in Appendix~\ref{app:singletPLODMR}, which studies the characteristics of the NV emission, only used the green laser---the IR laser was kept off.
All other experiments use both lasers, co-propagating through the fibre-taper.

An arbitrary waveform generator (AWG, Tektronix AWG70002A) is used to produce the MW signal sent to the antenna in addition to the reference signal used by the lock-in amplifier (Stanford Research Systems SR510).
The MW signal is amplified (Mini-circuits ZHL-25W-63+) and attenuated before being sent to the MW antenna. 
A data acquisition (DAQ) tool is used to measure the signals produced by the APD and the LIA.
For details on the design and characterization of the microwave antenna see Ref. \cite{Sorensen2025}.

\section{NV singlet photoluminescence ODMR}
\label{app:singletPLODMR}
\begin{figure}[b!]
    \centering
    \includegraphics[]{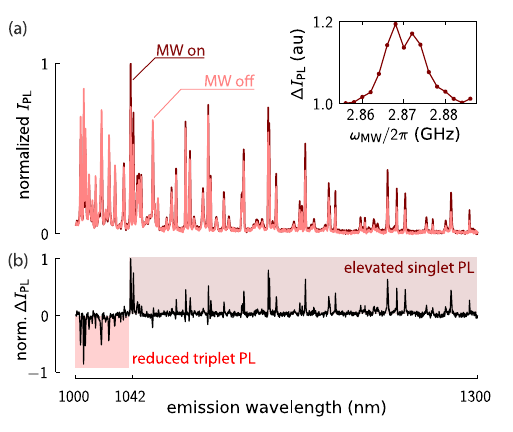}
    \caption{\textbf{Characterization of IR singlet cavity emission.} 
    {(a)}~Green excitation of the NV ensemble via the fibre-taper produces triplet ($^3\!E \rightarrow {}^3\!A_2$) and singlet ($^1\!A_1 \rightarrow {}^1\!E$) PL. The PL, whose intensity is spin dependent, couples into WGMs, then the fibre-taper.
    {(b)} 
    We plot the difference between the data shown in (a) and highlight regions that exhibit microwave-dependent triplet PL and singlet PL. 
    The inset in {(a)} shows zero-field singlet-PL ODMR measured using the cavity PL.
    }
    \label{fig:fig8_SingletPL}
\end{figure}
\begin{figure*}[b!]
    \centering
    \includegraphics[]{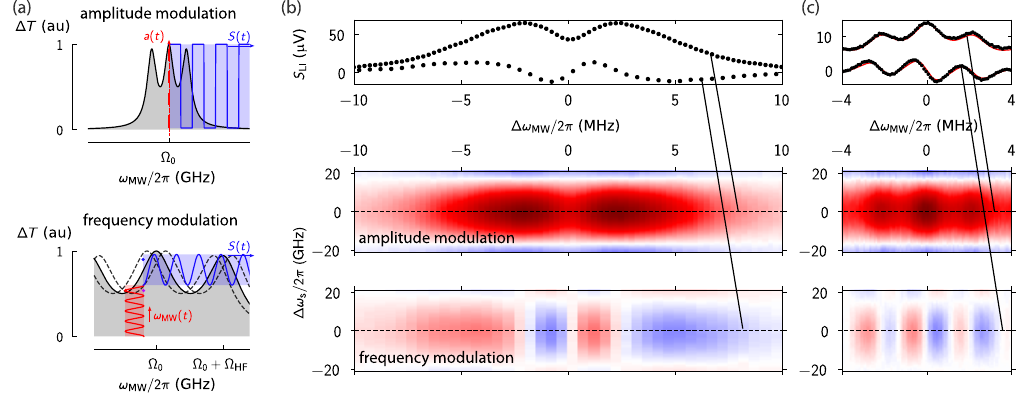}
    \caption{\textbf{Comparison of lock-in modulation techniques. } {(a)} We use two different modulation techniques: in one, the microwave amplitude, $a(t)$ is modulated on and off; in the other, the microwave frequency, $\omega_{\text{MW}}(t)$, is sinusoidally modulated about the centre frequency. 
    The first approach returns the total amplitude of the frequency-dependent change in transmission, whereas the second effectively takes the derivative of the frequency-dependent change in transmission.
    {(b)} For each technique we measure and plot (middle and bottom) the zero-field signal as a function of optical and microwave detuning, respectively $\Delta\omega_{\text{s}}$ and $\Delta\omega_{\text{MW}}$. 
    {(c)} We do the same for a single crystallographic orientation, detuned using a bias magnetic field.
    The amplitude of the signal produced by the first technique is larger than that produced by the second; however, the maximum contrast, maximum slope, and noise are improved using the second technique. 
    We therefore find the second technique to produce a better signal. 
    }
    \label{fig:fig9_LockInComparison}
\end{figure*}
In addition to performing singlet-absorption based magnetometry, we measured singlet-PL ODMR and inferred a magnetic field sensing sensitivity for this mechanism using our cavity. This Appendix describes these measurements.
Nanocavities modify the spectrum collected from an emitter both through cavity-enhanced emission and spectral filtering~\cite{Janitz2020,Masuda2024}.
To examine how the cavity alters the emissive characteristics of the NV ensemble, and therefore the sensitivity of singlet-PL magnetometry, we collect cavity-coupled PL using the dimpled fibre-taper [\Cref{fig:fig2_ODMR}~(a)]. 
Light from a 532~nm cw laser is scattered into the cavity from the fibre-taper, exciting the NV centres.
The NV centres emit both singlet and triplet PL into WGMs, which is subsequently coupled to the fibre-taper. 
Note that PL is only collected from WGMs, mitigating the collection of background PL from outside the microdisk.
The fibre-taper is positioned to maximize the collection rate of singlet PL (for more experimental details, see Appendix~\ref{app:SI-setup}).

Figure \ref{fig:fig8_SingletPL}~(a) compares the cavity emission spectrum with and without resonant microwave excitation of the zero-field spin transition.
Because the fibre-taper only collects light emitted into the cavity's WGMs, the measured spectrum consists of discrete cavity-filtered emission peaks.
While the singlet ZPL occurs near 1042~nm, we additionally observe a broad phonon sideband extending beyond 1300~nm, consistent with previous measurements of NV singlet emission~\cite{Marini2008}.
Note that we do not expect the singlet lifetime and subsequent rate of emission to be Purcell-enhanced by the cavity, as the singlet decay is dominated by non-radiative processes and broad emission into the PSB~\cite{Acosta2010-2}; however, similar to triplet PL, we do expect the cavity to induce some preferential emission into the WGMs~\cite{Masuda2024}.

Under resonant microwave excitation, the PL intensity above 1042~nm increases, while the visible and IR emission below 1042~nm decreases, consistent with enhanced singlet PL and reduced triplet PL, respectively [\Cref{fig:fig8_SingletPL}~(a)]. 
In WGMs near the ZPL, both triplet and singlet emission contribute to the observed PL.
The spin-dependent change in PL intensity, $\Delta I_{\text{PL}} = I_{\text{PL}}^{\text{MW on}} - I_{\text{PL}}^{\text{MW off}}$, is shown in \Cref{fig:fig8_SingletPL}~(b). 
Using this contrast mechanism, we perform zero-field singlet-PL ODMR, shown in the inset of \Cref{fig:fig8_SingletPL}~(a), and measure an ODMR contrast of approximately $20\,\%$.

In principle, singlet-PL ODMR could be used to perform magnetometry; however, singlet decay is dominated by non-radiative processes~\cite{Acosta2010-2}, resulting in low photon collection rates and correspondingly large photon-shot-noise compared to that of triplet-PL- and singlet-absorption ODMR.
In addition, phonon-assisted triplet PL extends into the spectral region surrounding the singlet ZPL, limiting the optical contrast. 
Consequently, despite the relatively strong ODMR contrast observed here as compared to triplet PL readout~\cite{Sorensen2025}, singlet-PL readout does not provide a competitive magnetic sensitivity in comparison to other readout modalities, such as triplet PL~\cite{Balasubramanian2008,Maze2008,Taylor2008}, singlet absorption~\cite{Acosta2010}, or spin-to-charge state conversion magnetometry\,\cite{Shields2015}. 
Using the rate of detection, we estimate the PSNL sensitivity based on singlet PL collected from the cavity to be $\sim10\,\upmu\text{T}\,\text{Hz}^{-1/2}$.
In this work, we therefore focus on singlet-absorption magnetometry.

\section{ODMR lock-in techniques}
\label{app:ODMRLITechniques}

The magnetic sensitivity can be improved using lock-in techniques, which reduce the impact of noise~\cite{El-Ella2017}.
In this work, we use two different lock-in schemes to measure the ODMR spectra of the NV ensemble, which each involve modulating the microwave drive field, whose time-dependent field amplitude is given by
\begin{align}
    A(t) & = a \cos\left(\omega_{\text{MW}} t \right).
\end{align}
In the first lock-in scheme, we modulate the amplitude of the MW field such that
\begin{align}
    a(t) & = a_0 \operatorname{square}(\omega_{\text{LO}} t).
\end{align}
Effectively, we turn the field on and off at the local oscillator frequency, $\omega_{\text{LO}}$, and lock in the resulting signal to that frequency. 
A sketch of this modulation scheme is shown at the top of \Cref{fig:fig9_LockInComparison}~(a), demonstrating how the microwave field amplitude modulation, denoted by the thickness of the red line, causes the fibre transmission to change. 
The lock-in amplifier returns the amplitude of the square time-modulated signal, $S(t)$.

In the second lock-in scheme, we modulate the frequency of the microwave drive field such that
\begin{align}
    \omega_{\text{MW}}(t) & = \bar{\omega}_{\text{MW}} + \delta \cos(\omega_{\text{LO}}t).
\end{align}
The amplitude of the frequency modulation, $\delta\sim 0.5$~MHz, is chosen to maximize the amplitude of the lock-in signal. 
A sketch of the second modulation scheme is shown at the bottom of \Cref{fig:fig9_LockInComparison}~(a), demonstrating how the microwave field frequency modulation causes the fibre transmission to vary in phase with the modulated field. 
The lock-in amplifier returns the amplitude of the time-modulated signal, $S(t)$. 
The second modulation technique requires a much smaller dynamic lock-in range, and, therefore, produces less noisy measurements.

We measure two ODMR spectra using each technique: first without a magnetic bias field [\Cref{fig:fig9_LockInComparison}~(b)], and second with a magnetic bias field. The latter permits us to measure the spectrum of a single NV orientation [\Cref{fig:fig9_LockInComparison}~(c)].
For each spectrum, we sweep both the microwave detuning and the laser detuning, producing a two-dimensional colour plot. 
The middle plots in \Cref{fig:fig9_LockInComparison}~(b,c) correspond to measurements that use the first amplitude-modulation schemes, and the bottom plots correspond to measurements that use the second frequency-modulation scheme. 
It is clear that while the first scheme produces measurements with larger signal amplitudes, the maximum response (slope) of the second technique is approximately two times better. 
Therefore, all magnetic sensitivity measurements use the second lock-in protocol.
Note that the modulation frequency of all measurements ($\omega_{\text{LO}}/2\pi \sim 20$~kHz) is much higher than the measurement bandwidth [$\sim 100$~Hz, as seen in \Cref{fig:fig6_measSens}~(b)].

\section{Experimental optimization of the magnetometer}
\label{app:optimization}
\begin{figure}[b!]
    \centering
    \includegraphics{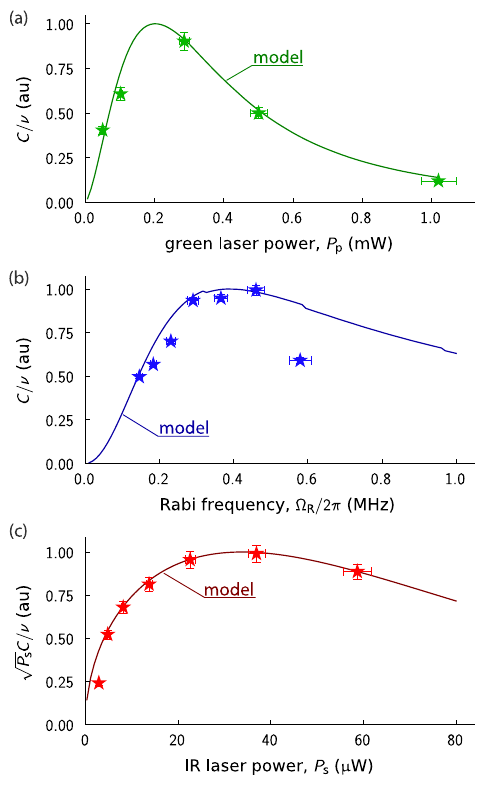}
    \caption{\textbf{Experimental optimization of the magnetometer.}  
    To optimize the ODMR, the ratio of contrast and linewidth, $C/\nu$, is maximized as a function of both green laser power (\textbf{a}) and microwave power, presented as Rabi frequency (\textbf{b}). 
    The Rabi frequency is extrapolated from the measurements done in Ref.~\cite{Sorensen2025}, for which there is significant uncertainty. 
    We also optimize the probe power (\textbf{c}) by maximizing the ratio of probe power and contrast to the linewidth, $\sqrt{P_{\text{s}}}C/\nu$.
    The contrast and linewidth are each extracted from the ODMR spectra obtained at each laser power and Rabi frequency. 
    Uncertainties in the values were estimated from the residuals of the covariance matrix obtained in the least-squares fit.
    The values predicted by the master-equation model [\Cref{app:MasterEquationModel}] are included as lineplots, which, alongside the data, have been normalized. 
    The green laser model data has also been scaled horizontally, given an inability to determine the calibration between intracavity green intensity and input power due to the unknown cavity coupling parameters of modes near the green pump. 
    }
    \label{fig:fig10_cwOptimization}
\end{figure}

Optimization of the optical and electronic environments [\Cref{eq:sensitivity}] involves maximizing the detected optical power, as well as the ratio of contrast and linewidth, $C/\nu$.
Both of these parameters are dependent on the optical intensity of the green and IR fields, as well as the microwave control power, each of which we experimentally tune to optimize the sensitivity. 
The results shown in \Cref{fig:fig10_cwOptimization} were used to optimize the experiment: we found that the optimized experiment used a cw green pump power $P_{\text{p}}\approx250\,\upmu$W, IR probe power $P_{\text{s}}\approx30\,\upmu$W and microwave power that produced a Rabi frequency $\Omega_{\text{R}}/2\pi\approx0.5\,$MHz. 
The model used to compare to the experimental results is detailed in \Cref{app:MasterEquationModel}.
We attribute the outlier in \Cref{fig:fig10_cwOptimization}~(b) to microwave heating and detuning of the cavity mode, which caused the contrast to appear lower than in reality. 

\section{Fitting the ODMR spectra}
\label{app:fits}

The parameters required to determine the PSNL sensitivity given by \Cref{eq:sensitivity} can be determined by fitting the relevant ODMR spectra, such as that seen in \Cref{fig:fig6_measSens}~(a). 
In this appendix, we discuss the different models used to fit the ODMR measurements and extract the relevant parameters.
For example, ODMR spectra acquired while driving at a single microwave frequency can be modelled using
\begin{align}
\begin{split}
    S^{(1)}(\Delta\omega_{\text{MW}}, C, \nu) = C[&\mathcal{L}(\Delta\omega_{\text{MW}}-\Omega_{\text{HF}}, \nu) \\
\quad&+ \mathcal{L}(\Delta\omega_{\text{MW}}, \nu) \\
\quad& + \mathcal{L}(\Delta\omega_{\text{MW}} + \Omega_{\text{HF}}, \nu)]\,,
    \end{split}
    \label{eq:fit_ODMR_1}
\end{align}
where $C$ is the signal amplitude and 
\begin{align}
    \mathcal{L}(\Delta, \nu) \coloneqq \frac{\left(\frac{\nu}{2}\right)^2}{\Delta^2 + \left(\frac{\nu}{2}\right)^2}\label{eq:fit_lorentzian}
\end{align}
is a Lorentzian centred at zero with a full-width at half-maximum (FWHM) equal to $\nu$.

\begin{figure*}[b!]
    \centering
    \includegraphics[]{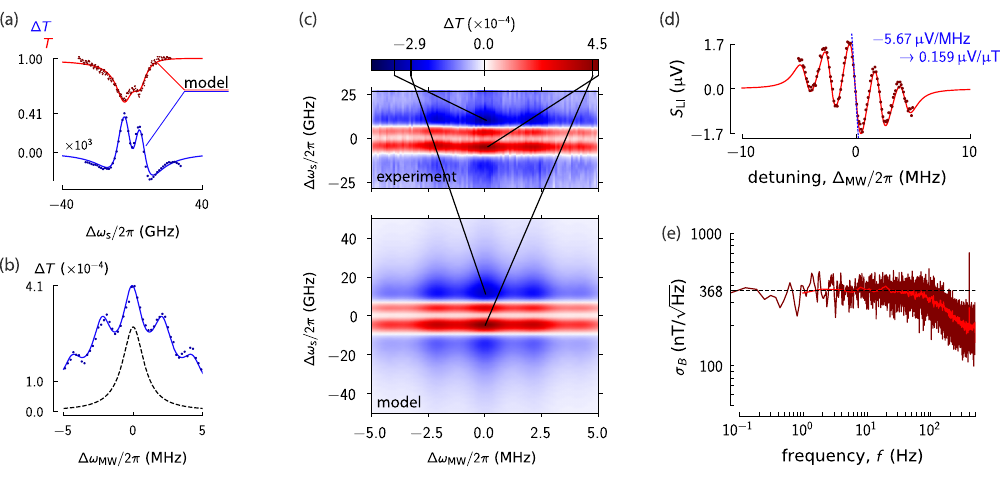}
    \caption{\textbf{Magnetic sensing using a higher Q mode on a different nanocavity. } 
    We use the model to predict the change in transmission as a function of optical {(a)} and microwave {(b)} detuning using the cavity parameters extracted from the fit to the experimental transmission data (a). {(c)}
    The model is used to calculate $\Delta T(\Delta\omega_{\text{MW}},\Delta\omega_{\text{s}})$ over a wide range and is in excellent agreement with that measured experimentally.
    {(d)} The measured ODMR spectrum reveals the point of highest sensitivity, whose slope is highlighted by the dashed blue line. 
    Driving the optimized spin ensemble at this point, we monitor the $S_{\text{LI}}(t)$ over 1000~s and convert it into a magnetic field signal. 
    {(e)} The power spectral density of the magnetic time trace gives the magnetic noise (dark red line). The light red line is a smoothed signal. 
    The mean of the magnetic noise in the flat region gives the sensitivity of the magnetometer.
    }
    \label{fig:fig11_highQ}
\end{figure*}

ODMR spectra produced while simultaneously driving all three hyperfine transitions, such as that shown in \Cref{fig:fig2_ODMR}~(d), are fit using
\begin{equation}
\begin{split}
    S^{(3)}(\Delta\omega_{\text{MW}}, C, \nu) = \frac{C}{3} \Big[ &\mathcal{L}(\Delta\omega_{\text{MW}}-2\Omega_{\text{HF}}, \nu)  \\
    \quad&+2\mathcal{L}(\Delta\omega_{\text{MW}}-\Omega_{\text{HF}}, \nu) \\
    \quad&+3\mathcal{L}(\Delta\omega_{\text{MW}}, \nu) \\   
    \quad&+2\mathcal{L}(\Delta\omega_{\text{MW}}  + \Omega_{\text{HF}}, \nu) \\
    \quad&+\mathcal{L}(\Delta\omega_{\text{MW}} + 2 \Omega_{\text{HF}}, \nu)\Big]\,.
\end{split}\label{eq:fit_ODMR_3}
\end{equation}
Equation \ref{eq:fit_ODMR_3} is true in the limit that $\Omega_{\text{R}}\ll \Omega_{\text{HF}}$, which approximately holds in this work.
The use of three drive tones increases the contrast of the central resonance dip, but manifests in two additional dips in the spectrum~\cite{Chakravarty2026}. 
These result from the convolution of the three hyperfine signatures with the three drive tones.
Following from \Cref{eq:fit_ODMR_1,eq:fit_ODMR_3}, the equation used to fit the secondary lock-in cw ODMR spectra seen in \Cref{fig:fig6_measSens}~(a) is given by
\begin{equation}
\begin{split}
    S^{(3)}_{\text{LI}}(\Delta\omega_{\text{MW}}, C, \nu, \delta) = &S^{(3)}(\Delta\omega_{\text{MW}}-\delta, C, \nu)  \\
    \quad&- S^{(3)}(\Delta\omega_{\text{MW}}+\delta, C, \nu)\,,\end{split}\label{eq:fit_ODMR_3_LI}
\end{equation}
where $\delta \sim 0.5$\,MHz is the amplitude of the microwave frequency modulation.

\section{Magnetic sensitivity of a high-$Q$ cavity}
\label{app:highQ}

\setlength{\extrarowheight}{2pt}
\begin{table*}[h!]
\centering
\caption{Experimental parameters for each of the cavity sensors investigated.  }
\begin{tabular*}{\linewidth}{@{\extracolsep{\fill}}lcccc}
parameter & symbol & low-$Q$ cavity value & high-$Q$ cavity value & unit \\ \hline \hline 
centre wavelength & $\lambda$     & 1040.9 & 1041.4 & nm \\
cavity quality$^{\dagger}$ & $Q_{\text{cav}}$     & $7.6\,(2)\times10^3$ & $35\,(3)\times10^3$ & --- \\
input power & $P_{\text{s}}$     & 18~(2) & $8.4\,(8)$ & $\upmu$W \\
detected power & $P_{\text{det}}$     & 0.42~(13) & 0.67~(20) & $\upmu$W \\
optimized  resonant cavity intensity & $I_{\text{s}}$     & 1.3~(4) & 2.3~(7) & MW/cm$^2$ \\
spin linewidth & $\nu$     & 1.82~(2) & 1.75~(3) & MHz \\
change in transmission & $\Delta T$     & $3.75\,(3)\times 10^{-4}$ & $4.15\,(7)\times10^{-4}$  & --- \\
contrast & $C$     & $4.15\,(4)\times 10^{-4}$ & $2.87\,(5)\times 10^{-4}$  & --- \\
sensitivity (measured) & $\eta$     & 253~(9) & 368~(12) & nT$\,\text{Hz}^{-1/2}$ \\
sensitivity (experimental PSNL) & $\eta_{\text{PSNL}}$     & 81~(12) & 89~(14) & nT$\,\text{Hz}^{-1/2}$ \\
sensitivity (model PSNL) & $\eta_{\text{PSNL}}^{\text{(model)}}$     & 66 & 87 & nT$\,\text{Hz}^{-1/2}$ \\ \hline \hline
\label{tab:tab3_cavityParameters}
\end{tabular*}
\begin{minipage}{\linewidth}
\raggedright
\small
$^{\dagger}$For the symmetric mode for the high-$Q$ cavity.
\end{minipage}
\end{table*}

To assess the effect of $Q$ on the magnetic sensitivity, we measure the magnetic sensitivity of a higher $Q$ WGM on a different microdisk cavity with a diameter of $\sim4.4\,\upmu$m.
The WGM is resonant at $1041.4$~nm and exhibits a doublet with resolved mode splitting between standing-wave modes formed by symmetric and anti-symmetric superpositions of clockwise and counter-clockwise travelling-wave modes~\cite{Coutts2026}, whose quality factors are $Q_{\text{cav}}^{\text{(s)}}=38\times10^3$ and $Q_{\text{cav}}^{\text{(a)}}=35\times10^3$, respectively. 
As was done with the lower-$Q_{\text{cav}}$ mode studied in the main text, we fit the experimentally measured transmission spectrum and use the cavity parameters to predict the spin-dependent change in transmission [\Cref{fig:fig11_highQ}~(a,b)].  
The two-dimensional data show excellent agreement between the model and experiment [\Cref{fig:fig11_highQ}~(c)], further demonstrating the fidelity of our model. 
Sitting on resonance with the higher-contrast mode, we measure the response to a detuned, modulated microwave field [\Cref{fig:fig11_highQ}~(d)], using the same frequency-modulation lock-in measurement scheme used for the first mode.
This lets us determine the point of maximum response.
Fixing the microwave detuning at this point, we record a time trace and convert it to a magnetic field signal, as was done before. 
The noise spectral density yields a magnetic sensitivity of 368~(12)~nT$\,\text{Hz}^{-1/2}$, which is higher than the PSNL sensitivity of 89~(14)~nT$\,\text{Hz}^{-1/2}$.
We summarize the cavity and experimental parameters, and the model-predicted sensitivity for each WGM cavity in \Cref{tab:tab3_cavityParameters}.

Although the device produced both larger $\Delta T \approx 4.1\times10^{-4}$ and a similar PSNL sensitivity [$\eta_{\text{PSNL}}=89\,(14)\,\text{nT}\,\text{Hz}^{-1/2}$] compared to the other cavity WGM (fibre coupling was worse), the measured sensitivity was only $\eta=368\,(12)\,$nT$\,{\text{Hz}}^{-1/2}$.
We attribute this behaviour to the increased susceptibility of higher $Q$ cavities to temperature fluctuations~\cite{Sansa2016,Roy2018}.

\section{NV population dynamics and IR absorption}
\label{app:MasterEquationModel}

\begin{figure*}[t!]
    \centering
    \includegraphics{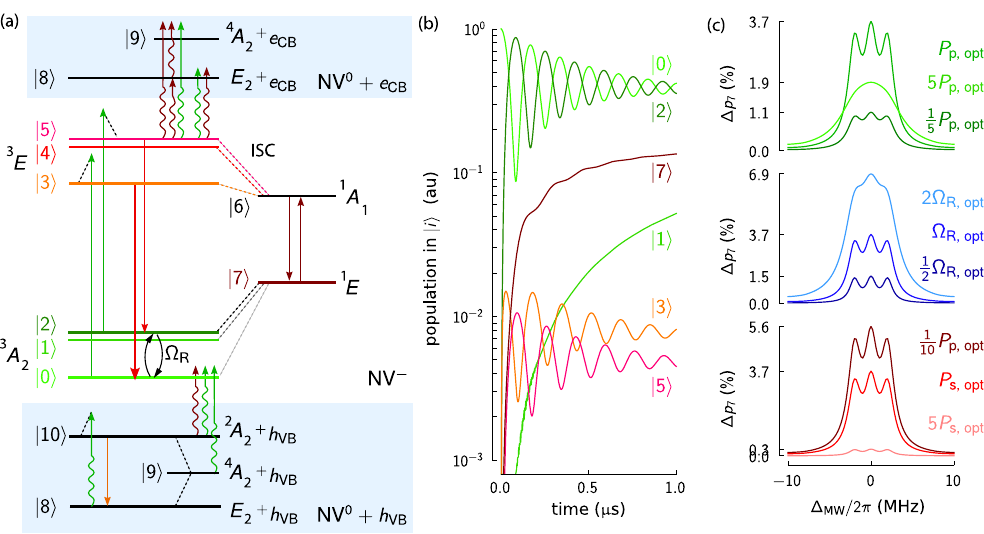}
    \caption{\textbf{Population dynamics of an NV centre}. {(a)} State diagram of the NV centre. Dashed lines indicate non-radiative decay processes, straight or wavy lines with up-facing (down-facing) arrows indicate incoherent driving (radiative decay) processes. Curved lines with arrows indicate the coherent microwave drive. See the text for discussion. 
    {(b)} A time-dependent solution to the Lindblad master equation for the eleven-level system in {(a)}. 
    In this example, $\Omega_{\text{R}}/2\pi = 6\,$MHz, $T_2^* = 0.43\,\upmu$s, $I_{\text{p}} = 10\,\text{kW/cm}^2$ and $I_{\text{s}} = 10\,\text{kW/cm}^2$. 
    The $|0\rangle\leftrightarrow|2\rangle$ spin transition is resonantly driven and Zeeman-detuned by 200\,MHz from the $|0\rangle\leftrightarrow|1\rangle$ transition. 
    Neither $I_{\text{p}}$, nor $I_{\text{s}}$ are large enough to saturate the $^3E$ or $^1A_1$ states.
    {(c)}
    We optimize the green and IR optical intensities and the Rabi flopping frequency to minimize the corresponding magnetic sensitivity (\Cref{eq:sensitivity}).
    The spin-dependent difference in $|7\rangle$ is plotted for a range of $I_{\text{p}}$, $I_{\text{s}}$, and $\Omega_{\text{R}}$. 
    Note that while ionization from $^1\!A_1$ is possible, its lifetime is so short that the transition can be ignored for the range of intensities investigated~\cite{Shandilya2025}. 
    Strictly, the transition $|9\rangle\rightarrow|0,1,2\rangle$ is possible by the absorption of two IR photons, but the transition rate is negligible~\cite{Shandilya2025} so we ignore it.
    }
    \label{fig:fig12_NVStateDiagram}
\end{figure*}
To predict the magnetic sensitivity given a range of environmental parameters, we solve the Lindblad master equation to obtain the time evolution and steady-state solutions to the density matrix for a single NV centre. 
This allows us to determine the optimal optical and microwave field strengths that maximize the spin-dependent change in population of the $^1\! E $ state. 
This, in turn, determines the optimal absorption contrast.
By incorporating the Lindblad superoperators of each electronic transition, our approach accounts for dissipation, dephasing, and coherent microwave driving of the ground-state spin-transition, in addition to optical pumping between electronic states of the NV centre.

Coherent dynamics of the NV centre are modelled using the Hamiltonian of the ground state spin triplet in the presence of a magnetic field $\mathbf{B}$~\cite{Gruber1997,Doherty2013}:
\begin{align}
    \hat{H}= D S_z^2 + \gamma_e \,\mathbf{B} \cdot \mathbf{S}, \label{eq:Hmag}
\end{align}
$D = 2.87$\,GHz is the zero-field ground state splitting parameter, and $\mathbf{S} = (S_x, S_y, S_z)$ is the electronic spin-1 operator.
Equation \ref{eq:Hmag} omits contributions from local electric fields, strain fields, and hyperfine nuclear interactions, which are suppressed by $D$ and by the presence of a bias magnetic field~\cite{Doherty2013,Barry2020}. 

Incoherent dynamics of the NV centre are modelled by incorporating the Hamiltonian with the jump operators $\{L_{i,j}\}$ and their corresponding transition rates $\gamma_{i,j}$, alongside the Lindblad master equation for the system's density matrix, $\hat{\rho}$:
\begin{align}
\begin{aligned}
    \frac{d \hat{\rho}}{d t}&=\mathcal{L} \hat{\rho} \\
    &=-i[\hat{H}, \hat{\rho}]+\sum_{i\neq j} \gamma_{i,j}\left[L_{i,j} \hat{\rho} L_{i,j}^{\dagger}-\frac{1}{2}\left\{L_{i,j}^{\dagger} L_{i,j}, \hat{\rho}\right\}\right].
    \label{eq:lindblad}
\end{aligned}
\end{align}
Spin dephasing and decoherence from the presence of local electronic and nuclear spin baths, strain and magnetic field gradients, and electric field noise are incorporated via the experimentally measured inhomogeneous dephasing time, $T_2^*$.

Specifically, we find steady-state solutions to \Cref{eq:lindblad} for the eleven-state system defined in \Cref{fig:fig12_NVStateDiagram}~(a) with environment-dependent transition rates. 
The transition rate of electronically permissible decay paths is given by the relevant phenomenological decay constants, and some state transitions are achieved by the absorption of one or two green or IR photons, the intensity-dependent rates of which are extracted phenomenologically~\cite{Shandilya2025}.
Note that we assume that the intensity-dependent transition rates in the presence of a 1042~nm field are identical to those of a 966~nm field studied in Ref.~\cite{Shandilya2025}.
For an axial magnetic field $B$, we rewrite \Cref{eq:Hmag} to describe coherent interactions between the classical microwave field and the ground state triplet:
\begin{equation}
\begin{aligned}
    \hat{H} &= \frac{\Omega_{\text{R}}}{2}\left( |1\rangle \langle 0| + |2\rangle \langle 0| + |0\rangle \langle 1| + |0\rangle \langle 2| \right) \\ 
    &\qquad+ (\Delta\omega_{\text{MW}}+\gamma_e B) |1\rangle \langle 1| + (\Delta\omega_{\text{MW}}-\gamma_e B) |2\rangle \langle 2|.
\end{aligned}
\end{equation}
Then, we summarize the jump operators and their corresponding transition rates in \Cref{tab:tabMEDecayRates,tab:tab5_MEExcitation}.

\setlength{\extrarowheight}{2pt}
\begin{table}[t!]
\centering
\caption{Summary of decay and dephasing rates between states of the NV centre. The transition rates are given in \Cref{tab:tab6_modelParams}. }
\begin{tabular*}{\linewidth}{@{\extracolsep{\fill}}ccc}
jump operator & matrix element        & transition rate          \\ \hline \hline
              & radiative &                            \\
$L_{3,0}$     & $|0\rangle\langle 3|$ & $k_{3,0}$ \\
$L_{4,1}$     & $|1\rangle\langle 4|$ & $k_{4,1}$ \\
$L_{5,2}$     & $|2\rangle\langle 5|$ & $k_{5,2}$ \\
$L_{6,7}$     & $|7\rangle\langle 6|$ & $k_{6,7}$ \\
$L_{10,8}$     & $|8\rangle\langle 10|$ & $k_{10,8}$ \\
              &                       &                            \\
              & non-radiative &                            \\
$L_{3,6}$     & $|6\rangle\langle 3|$ & $k_{3,6}$ \\
$L_{4,6}$     & $|6\rangle\langle 4|$ & $k_{4,6}$ \\
$L_{5,6}$     & $|6\rangle\langle 5|$ & $k_{5,6}$ \\
$L_{7,0}$     & $|0\rangle\langle 7|$ & $k_{7,0}$ \\
$L_{7,1}$     & $|1\rangle\langle 7|$ & $k_{7,1}$ \\
$L_{7,2}$     & $|2\rangle\langle 7|$ & $k_{7,2}$ \\
$L_{10,9}$     & $|9\rangle\langle 10|$ & $k_{10,9}$ \\
$L_{9,8}$     & $|8\rangle\langle 9|$ & $k_{9,8}$ \\
 & & \\
              & dephasing &                            \\
$L_{1,1}$     & $|1\rangle\langle 1|$ & $2\gamma_{\phi}$ \\
$L_{2,2}$     & $|2\rangle\langle 2|$ & $2\gamma_{\phi}$ \\
$L_{4,4}$     & $|4\rangle\langle 4|$ & $2\gamma_{\phi}$ \\
$L_{5,5}$     & $|5\rangle\langle 5|$ & $2\gamma_{\phi}$ \\ \hline \hline
\label{tab:tabMEDecayRates}
\end{tabular*}
\end{table}

We first assume that the entire population starts in $|0\rangle$, corresponding to the $|m_{\text{s}} = 0\rangle$ ground state of the negatively charged NV centre. 
A cw microwave field coherently drives the ground state population between the triplet $|0\rangle$ and $|1,2\rangle$ spin states at angular frequency $\Omega_{\text{R}}$, and a green pump field (with excitation rates $\Gamma^{\text{p}}_{i,j} = \sigma^{\text{p}}_{i,j} I_{\text{p}} /\hbar \omega_{\text{p}}$, $i=0,1,2$, and $j=3,4,5$) incoherently drives the ground states ($^3\!A_2$, $|0,1,2\rangle$) into the excited states ($^3\!E$, $|3,4,5\rangle$), conserving spin. 
Population of the excited states primarily decays through two mechanisms: direct, spin-conserving radiative decay ($^3\!E\rightarrow\, ^{3}\!A_2$), or indirectly through the singlet state via the intersystem crossing (ISC) ($^3\!E\rightarrow\,^{1}\!A_1\rightarrow\,^1E\rightarrow\, ^{3}\!A_2$). 
Transition rates through the ISC are spin-state-dependent ($k_{36}<k_{46}, k_{56}$), creating a mechanism through which the NV spin state can be determined. 

In \Cref{fig:fig12_NVStateDiagram}~(b) we plot the population density of several NV electronic states in the presence of a microwave drive field, a green optical field, and a 1042\,nm field as predicted by the described master equation model.
The strong resonant microwave field is shown to cause the ground state NV centre to Rabi flop $(|0\rangle \leftrightarrow |2\rangle)$, and, notably, the metastable state ($|7\rangle$) is non-negligibly populated due to its long lifetime. 
We then calculate the steady-state population across a range of experimental parameters ($\Omega_{\text{R}},\,I_{\text{p}},\,I_{\text{s}}$) and find the optimal set ($\Omega_{\text{R,opt}},\,I_{\text{p,opt}},\,I_{\text{s,opt}}$) which maximizes the expression $\sqrt{I_{\text{s}}}\Delta p_7/\nu$, where $\Delta p_7 = p_7^{\text{(on)}}-p_7^{\text{(off)}}$ is the resonant spin-dependent change in population of $|7\rangle$ and determines the optical contrast, and $\sqrt{I_{\text{s}}}$ estimates the contribution from photon-shot-noise.

Figure \ref{fig:fig12_NVStateDiagram}~(c) demonstrates the effect of each experimental parameter on the resulting ODMR profile, as predicted by the master-equation model. 
In each panel, we vary one experimental parameter at a time. 
For high $I_{\text{p}}$ and $\Omega_{\text{R}}$ we observe power broadening of the spin linewidth, reducing the signal response. 
The model predicts that at the optimal set of experimental parameters (\Cref{tab:tab7_OptParametersMain}), the single-NV spin-dependent population has a contrast of $\Delta p_7^{(\text{single})} = 2.72\,\%$.

When driving a single NV crystallographic orientation in an ensemble, the absorption contrast is reduced by a factor of four because only a quarter of the total NV centres are resonantly driven. 
Furthermore, singlet absorption is polarization dependent. 
Therefore, the average absorption cross-section of an ensemble is scaled by a factor $\xi$; for a $(100)$ cut diamond, $\xi = 1/3$ for a field polarized in (TE) or out of (TM) plane, which is a good approximation for the cavity WGMs. 
This both reduces the baseline singlet absorption, as well as the change in absorption for a single NV crystallographic orientation. 
Taking these effects into consideration, we use \Cref{eq:absCoefficient} to calculate the microwave-dependent absorption coefficient of the diamond cavity ($\alpha^{(\text{off})}$ and $\alpha^{(\text{on})}$) and the nominal change in the absorption coefficient ($\Delta\alpha$, \Cref{tab:tab7_OptParameters}) at the optimal environmental conditions. 
The relative change in absorption sets the contrast and, therefore, the minimum PSNL magnetic sensitivity for a singlet-absorption magnetometer.
For an IR field polarized along the NV axis, $\Delta\alpha$ will be larger ($\xi = 1$). 

\setlength{\extrarowheight}{2pt}
\begin{table*}[b!]
\centering
\caption{Summary of excitation rates between states of the NV centre. The absorption cross-section values are given in \Cref{tab:tab6_modelParams}.}
\begin{tabular}{ccc}
jump operator & matrix element        & excitation rate, $\gamma$            \\ \hline \hline
$L_{0,3}$     & $|3\rangle\langle 0|$ & $\sigma^{\text{p}}_{0,3}I_{\text{p}}/\hbar\omega_{\text{p}}$ \\
$L_{1,4}$     & $|4\rangle\langle 1|$ & $\sigma^{\text{p}}_{1,4}I_{\text{p}}/\hbar\omega_{\text{p}}$ \\
$L_{2,5}$     & $|5\rangle\langle 2|$ & $\sigma^{\text{p}}_{2,5}I_{\text{p}}/\hbar\omega_{\text{p}}$ \\
$L_{8,10}$     & $|10\rangle\langle 8|$ & $\sigma_{\text{8,10}}^{\text{p}} I_{\text{p}}/\hbar\omega_{\text{p}}$ \\
$L_{7,6}$     & $|6\rangle\langle 7|$ & $\sigma^{\text{s}}_{7,6}I_{\text{s}}/\hbar\omega_{\text{s}}$ \\
$L_{3,8}$     & $|8\rangle\langle 3|$ & $\sigma_{3,8}^{\text{p}}I_{\text{p}}/\hbar\omega_{\text{p}} + \sigma_{3,8}^{\text{s},1}I_{\text{s}}/\hbar\omega_{\text{s}}$ \\
$L_{4,8}$     & $|8\rangle\langle 4|$ & $\sigma_{4,8}^{\text{p}}I_{\text{p}}/\hbar\omega_{\text{p}} + \sigma_{4,8}^{\text{s},1}I_{\text{s}}/\hbar\omega_{\text{s}}$ \\
$L_{5,8}$     & $|8\rangle\langle 5|$ & $\sigma_{5,8}^{\text{p}}I_{\text{p}}/\hbar\omega_{\text{p}} + \sigma_{5,8}^{\text{s},1}I_{\text{s}}/\hbar\omega_{\text{s}}$ \\
$L_{3,9}$     & $|9\rangle\langle 3|$ & $\sigma_{3,9}^{\text{p}}I_{\text{p}}/\hbar\omega_{\text{p}} + \left(\sigma_{3,9}^{\text{s},1}I_{\text{s}} + \sigma_{3,9}^{\text{s}, 2}I_{\text{s}}^2\right)/\hbar\omega_{\text{s}}$ \\
$L_{4,9}$     & $|9\rangle\langle 4|$ & $\sigma_{4,9}^{\text{p}}I_{\text{p}}/\hbar\omega_{\text{p}} + \left(\sigma_{4,9}^{\text{s},1}I_{\text{s}} + \sigma_{4,9}^{\text{s}, 2}I_{\text{s}}^2\right)/\hbar\omega_{\text{s}}$ \\
$L_{5,9}$     & $|9\rangle\langle 5|$ & $\sigma_{5,9}^{\text{p}}I_{\text{p}}/\hbar\omega_{\text{p}} + \left(\sigma_{5,9}^{\text{s},1}I_{\text{s}} + \sigma_{5,9}^{\text{s}, 2}I_{\text{s}}^2\right)/\hbar\omega_{\text{s}}$ \\
$L_{9,0}$     & $|0\rangle\langle 9|$ & $\sigma_{9,0}^{\text{p}}I_{\text{p}}/\hbar\omega_{\text{p}}$ \\
$L_{9,1}$     & $|1\rangle\langle 9|$ & $\sigma_{9,1}^{\text{p}}I_{\text{p}}/\hbar\omega_{\text{p}}$ \\
$L_{9,2}$     & $|2\rangle\langle 9|$ & $\sigma_{9,2}^{\text{p}}I_{\text{p}}/\hbar\omega_{\text{p}}$ \\
$L_{10,0}$     & $|0\rangle\langle 10|$ & $\sigma_{10,0}^{\text{p}}I_{\text{p}}/\hbar\omega_{\text{p}} + \sigma_{10,0}^{\text{s,1}}I_{\text{s}}/\hbar\omega_{\text{s}}$ \\
$L_{10,1}$     & $|1\rangle\langle 10|$ & $\sigma_{10,1}^{\text{p}}I_{\text{p}}/\hbar\omega_{\text{p}} + \sigma_{10,1}^{\text{s,1}}I_{\text{s}}/\hbar\omega_{\text{s}}$ \\
$L_{10,2}$     & $|2\rangle\langle 10|$ & $\sigma_{10,2}^{\text{p}}I_{\text{p}}/\hbar\omega_{\text{p}} + \sigma_{10,2}^{\text{s,1}}I_{\text{s}}/\hbar\omega_{\text{s}}$ \\ \hline \hline
\label{tab:tab5_MEExcitation}
\end{tabular}
\end{table*}

\setlength{\extrarowheight}{1.25pt}
\begin{table*}[t!]  
\centering 
\caption{NV-related parameters and constants. }  
\begin{tabular*}{\linewidth}{@{\extracolsep{\fill}}l l c c}   
\hline\hline   
Parameter  & Description & Value & Reference \\ 
\hline\hline  
$k_{3,0}, k_{4,1}, k_{5,2}$  &  transition rate & 66\,(5)\,MHz & \cite{Tetienne2012} \\
$k_{3,6}$  &  transition rate & 8\,(4)\,MHz & \cite{Tetienne2012} \\
$k_{4,6}, k_{5,6}$  &  transition rate & 53\,(7)\,MHz & \cite{Tetienne2012} \\
$k_{6,7}$  &  transition rate & 1000\,MHz & \cite{Tetienne2012} \\
$k_{7,0}$  &  transition rate & 1.0\,(8)\,MHz & \cite{Tetienne2012} \\
$k_{7,1}$, $k_{7,2}$  &  transition rate & 0.7\,(5)\,MHz & \cite{Tetienne2012} \\
$k_{10,8}$  &  transition rate & 53\,MHz & \cite{Tetienne2012} \\
$k_{10,9}$  &  transition rate & 4.7\,kHz & \cite{Shandilya2025} \\
$k_{9,8}$  &  transition rate & 39\,MHz & \cite{Shandilya2025} \\
$\Omega_{\text{R}}$  & MW Rabi frequency & ---- & ---- \\
$\sigma^{\text{p}}_{0,3} = \sigma^{\text{p}}_{1,4} = \sigma^{\text{p}}_{2,5}$  & green triplet absorption cross-section & $3.1\,(8)\times10^{-21}\,\text{m}^2$ & \cite{Wee2007,Fraczek2017} \\
$\sigma^{\text{s}}_{7,6}$  & IR singlet absorption cross-section & $2.0\,(3)\times10^{-22}\,\text{m}^2$ & \cite{Dumeige2013} \\
$\sigma_{\text{3,8}}^{\text{p}}=\sigma_{\text{4,8}}^{\text{p}}=\sigma_{\text{5,8}}^{\text{p}}$  & green pump absorption cross-section & $7.2\times 10^{-26}\,\text{m}^2$  & \cite{Shandilya2025}  \\
$\sigma_{\text{3,8}}^{\text{s,1}}=\sigma_{\text{4,8}}^{\text{s,1}}=\sigma_{\text{5,8}}^{\text{s,1}}$  & IR probe absorption cross-section---one photon & $1.5\times 10^{-24}\,\text{m}^2$ & \cite{Shandilya2025} \\
$\sigma_{\text{3,9}}^{\text{p}}=\sigma_{\text{4,9}}^{\text{p}}=\sigma_{\text{5,9}}^{\text{p}}$  & green pump absorption cross-section & $6.0\times 10^{-24}\,\text{m}^2$ & \cite{Shandilya2025} \\
$\sigma_{\text{3,9}}^{\text{s,1}}=\sigma_{\text{4,9}}^{\text{s,1}}=\sigma_{\text{5,9}}^{\text{s,1}}$  & IR probe absorption cross-section---one photon & $7.2\times 10^{-24}\,\text{m}^2$ &  \cite{Shandilya2025}\\
$\sigma_{\text{3,9}}^{\text{s,2}}=\sigma_{\text{4,9}}^{\text{s,2}}=\sigma_{\text{5,9}}^{\text{s,2}}$  & IR probe absorption cross-section---two photon & $5.4\times 10^{-52}\,\text{m}^4$ & \cite{Shandilya2025} \\
$\sigma_{\text{9,0}}^{\text{p}}=\sigma_{\text{9,1}}^{\text{p}}=\sigma_{\text{9,2}}^{\text{p}}$  & green pump absorption cross-section & $4.0\times 10^{-22}\,\text{m}^2$ & \cite{Shandilya2025} \\
$\sigma_{\text{10,0}}^{\text{s,1}}=\sigma_{\text{10,1}}^{\text{s,1}}=\sigma_{\text{10,2}}^{\text{s,1}}$  & IR probe absorption cross-section---one photon & $3.8\times 10^{-24}\,\text{m}^2$ & \cite{Shandilya2025} \\
$\sigma_{\text{10,0}}^{\text{p}}=\sigma_{\text{10,1}}^{\text{p}}=\sigma_{\text{10,2}}^{\text{p}}$  & green pump absorption cross-section & $1.3\times 10^{-25}\,\text{m}^2$ & \cite{Shandilya2025} \\
$\sigma_{\text{8,10}}^{\text{p}}$  & green pump absorption cross-section & $1.7\,(1)\times10^{-21}\,\text{m}^2$ & \cite{Fraczek2017} \\
$I_{\text{p}}$  & green pump average intensity & ---- & ---- \\
$I_{\text{s}}$  & IR probe average intensity & ---- & ---- \\
$\lambda_{\text{p}}$  & green pump wavelength & 532\,nm & ---- \\
$\lambda_{\text{s}}$  & IR probe wavelength & ---- & ---- \\
$\lambda_{\text{ZPL}}$  & singlet ZPL wavelength & $\sim$1042\,nm & ---- \\
$\omega_{\text{p}}/2\pi$  & green pump frequency & 563\,THz & ---- \\
$\omega_{\text{s}}/2\pi$  & IR probe frequency & ---- & ---- \\
$\omega_{\text{ZPL}}/2\pi$  &  singlet ZPL frequency & $\sim$288\,THz & ---- \\
$\omega_{\text{cav}}/2\pi$  & WGM resonant frequency & ---- & ---- \\
$T_2^*$  & inhomogeneous dephasing time & $0.43\,\upmu$s & \cite{Sorensen2025} \\
$\gamma_{\phi}\approx (T_2^*)^{-1}$  & dephasing rate & $2.33$\,MHz & ---- \\
$M$  & NV density$^\dagger$ & $7.9\times10^{23}\,\text{m}^{-3}$ or $4.5$\,ppm & \cite{ElementSix2021} \\    
\hline 
\hline
\end{tabular*} 
\label{tab:tab6_modelParams}
\begin{minipage}{\linewidth}
\raggedright
\small
$^\dagger$$M$ is the total NV density and is not scaled for the density of a single NV orientation.
\end{minipage}
\end{table*} 

\clearpage

\setlength{\extrarowheight}{2pt}
\begin{table}[t!]
\centering
\caption{Microwave-drive-dependent steady state  population and IR absorption contrast for a field polarized in or out of plane of a $(100)$ cut diamond. }
\begin{tabular}{ccc}
parameter & value & unit \\ \hline \hline 
$p_7^{(\text{on, single})}$ & 11.43 & \% \\
$p_7^{(\text{off, single})}$ & 8.71 & \% \\
$\Delta p_7^{(\text{single})}$ & 2.72 & \% \\
$\alpha^{(\text{on})}$ & $4.95^{\dagger}$ & m$^{-1}$\\
$\alpha^{(\text{off})}$ & 4.59 & m$^{-1}$ \\
$\Delta\alpha$ & 0.36 & m$^{-1}$ \\ \hline \hline
\label{tab:tab7_OptParameters}
\end{tabular}
\begin{minipage}{\linewidth}
\raggedright
\small
$^\dagger$Pertains to a single driven NV orientation of an ensemble.
\end{minipage}
\end{table}

\section{Analytical comparison of singlet-absorption magnetometry approaches}
\label{app:comparison}
Here, we derive the magnetic sensitivity of singlet-absorption magnetometers based on a single-pass scheme, a Fabry--Perot etalon, and a side-coupled WGM cavity to quantify the benefit of cavity integration.
Throughout, we neglect detector efficiency and assume that the only optical loss is absorption by the NV singlet transition, with the field polarized along $[100], [010], $ or $[001]$, so that the absorption coefficients from the previous section apply directly. 
There is, of course, absorption loss associated with other defects in the diamond~\cite{Coutts2026}, which is lumped into the intrinsic loss term ($\kappa_{\text{i}}$). Therefore, the sensitivity limits here are highly idealized. Our methods, however, allow for sensitivity limits for non-ideal cavities to be determined if the other sources of loss are known.

\subsection{Single-pass absorption}
Consider a coherent, spatially homogeneous optical field with cross-sectional area $A$ and wavelength $\lambda_{\text{ZPL}}$ matching the NV singlet transition passing through a diamond sample of length $L$.
Defining the average absorption coefficient as the average of the two spin-dependent absorption coefficients [$\alpha_0 \equiv (\alpha^{\text{(on)}} + \alpha^{\text{(off)}})/2$] and the spin-dependent change in absorption ($\Delta\alpha \equiv \alpha^{\text{(on)}} - \alpha^{\text{(off)}}$), the transmission is (neglecting Fresnel losses)
\begin{align}
    T^{(j)}_{\text{single}}(L) = \exp(-\alpha^{(j)}L),
    \label{eq:TBulk}
\end{align}
where $j \in \{\text{on},\,\text{off}\}$. 
This gives spin-readout contrast
\begin{align}
    C =  \frac{\left|T^{(\text{on})} - T^{(\text{off})}\right|}{T_0} \,,
    \label{eq:CBulk}
\end{align}
where $T_0 = (T^{(\text{on})} + T^{(\text{off})})/2$ is the average transmission. 
To optimize the magnetic sensitivity, the optical intensity should match that determined in the previous section ($I_{\text{s,opt}}$). 
Therefore, the average detected power is 
\begin{align}
P_{\text{det}} =I_{\text{s,opt}}\, A\, T_0.
\label{eq:PBulk}
\end{align}
Combining \Cref{eq:TBulk,eq:CBulk,eq:PBulk} with \Cref{eq:sensitivity}, we find that the sensitivity is given by
\begin{align}
    \eta_{\text{single}} = \beta\frac{\nu}{\sqrt{I_{\text{s,opt}}\,A}}\frac{{\exp\left(\frac{\alpha_0\,L}{2}\right)}}{2\sinh\left(\frac{\Delta \alpha\, L}{2}\right)}.
    \label{eq:etacwBulk}
\end{align}
Strictly, $\nu$ depends on the optical intensity.
Therefore $\nu$ varies with $L$ as the transmitted optical field passes through the absorbing material; this dependence is weak, however, and we take $\nu$ to be constant for simplicity.
In the small-sensor limit ($\alpha_0\, L\ll 1$ and $\Delta\alpha\, L\ll 1$) we find that 
\begin{align}
    \eta_{\text{single}} \approx \beta \frac{\nu}{\sqrt{I_{\text{s,opt}}\,A}}\frac{1}{\Delta \alpha\,L}.
    \label{eq:etacwBulkSimple}
\end{align}

In the small-sensor limit, the sensitivity improves as $1/L$; however, as $L$ approaches the optical penetration depth of the diamond, the detected intensity decreases significantly while the improvement in spin contrast saturates, causing the sensitivity to level off and eventually worsen.
Note also that both $\alpha_0$ and $\Delta\alpha$ depend on intensity and are therefore also dependent on $L$. 
This effect only becomes significant when $\alpha_0\,L\sim 1$; the sensitivities calculated using both constant and intensity-dependent absorption coefficients remain similar over the experimentally relevant range and diverge only when $L \gg \alpha_0^{-1}$.

\subsection{Fabry--Perot cavity absorption}

We now compare the sensitivity of the single-pass scheme to that of an etalon, created by adding two identical mirrors to the diamond sample described above. 
Following the standard analysis of an etalon~\cite{Dumeige2013}, the cavity power transmission on resonance is given by
\begin{align}
    T^{({j})} = \frac{T_{\text{m}}^2 \exp({-\alpha^{({j})} L})}{\left[1 - (1-T_{\text{m}})\exp({-\alpha^{{j})} L})\right]^2},
    \label{eq:TFPCav}
\end{align}
where $T_{\text{m}}$ ($R_{\text{m}} = 1 - T_{\text{m}}$) is the mirror power transmittance (reflectance) and $j\in\{\text{on}, \text{off}\}$.
The contrast can be calculated using \Cref{eq:CBulk} alongside \Cref{eq:TFPCav}, and we choose the input power such that the intracavity intensity equals $I_{\text{s,opt}}$. 
In this case it can be shown that the detected power is simply given by:
\begin{align}
    P_{\text{det}} = I_{\text{s,opt}}\,A\,T_{\text{m}}\exp(-\alpha_0\,L).
    \label{eq:PdetFPcav}
\end{align}

Combining \Cref{eq:sensitivity,eq:CBulk,eq:TFPCav,eq:PdetFPcav} we obtain the cw PSNL sensitivity of a diamond sensor in an etalon:
\begin{align}
\begin{aligned}
    \eta_{\text{etalon}} &= \frac{\beta}{\sqrt{T_{\text{m}}}}\frac{\nu}{\sqrt{I_{\text{s,opt}}\,A}}\frac{\exp\left(\frac{\alpha_0\,L}{2}\right)}{2\sinh\left(\frac{\Delta\alpha\,L}{2}\right)} 
    \\&\times \frac{\left[1-T_{\text{m}}-\exp\left(\alpha^{\text{(off)}}\,L\right)\right]^2\left[1-T_{\text{m}}-\exp\left(\alpha^{\text{(on)}}\,L\right)\right]^2}{\left[1-T_{\text{m}}+\exp(\alpha_0\,L)\right]\left[T_{\text{m}}-1+\exp(\alpha_0\,L)\right]^3}.
\end{aligned}
    \label{eq:etaFPcav}
\end{align}
For $T_{\text{m}}=1$ in the small-sensor limit, \Cref{eq:etaFPcav} reduces to an expression identical to \Cref{eq:etacwBulkSimple} as expected.
Furthermore, in the small-sensor limit  we find that \Cref{eq:etaFPcav} has a minimum when $T_{\text{m}} \approx \alpha_0\,L$. 
Hence, in the small-sensor limit
\begin{align}
    \eta_{\text{etalon}} \approx \beta\frac{\nu}{\sqrt{I_{\text{s,opt}}\,A}}\frac{\sqrt{\alpha_0}}{\Delta\alpha\sqrt{L}}.
    \label{eq:etaFPcavSimp}
\end{align}
For sensors outside the small-sensor limit we numerically evaluate \Cref{eq:etaFPcav} to find the optimal $T_{\text{m}}$.

Comparing \Cref{eq:etacwBulkSimple,eq:etaFPcavSimp} we find that, in the small-sensor limit, an etalon improves the magnetic sensitivity by a factor of 
\begin{align}
    \frac{\eta_{\text{single}}}{\eta_{\text{etalon}}} = \frac{1}{\sqrt{\alpha_0\,L}}. 
\end{align}

In the case that $L\gtrsim1/\alpha_0$ we find that the optimal $T_{\text{m}}\rightarrow1$, corresponding to negligible cavity enhancement; in this case, the sensitivity approaches that of the single-pass sensor [\Cref{fig:fig4_model}\,(c)].

\subsection{Side-coupled cavity absorption}
Now, we consider the sensitivity of a side-coupled cavity like the one experimentally studied in this work. 
A coherent optical field travels through a waveguide, which is side-coupled to a WGM cavity with a resonance frequency equal to the singlet ZPL $\omega_{\text{ZPL}}$. 
The cavity transmission for a singlet mode is given by~\cite{Coutts2026}
\begin{align}
    T^{({j})} =  &\left| \tau - \frac{2\kappa_{\text{ex}}}{\kappa_{\text{ex}} + \kappa_{\text{i}}+ \kappa_{\text{p}} + \kappa_{\text{abs}}^{{(j)}}-i\Delta\omega_{\text{s}}}\right|^2
    \label{eq:Tcav}\,.
\end{align}
The expression in the case of a WGM cavity supporting a doublet mode can be found in Ref.~\cite{Coutts2026}.
Here, $j\in\{\text{on}, \text{off}\}$, $\tau$ is a complex number describing the fibre-taper transmission, which is used to describe Fano interference effects, and $\kappa_{\text{abs}}^{(\text{on})}$ and $\kappa_{\text{abs}}^{(\text{off})}$ are the absorption loss rates inside the cavity with the microwave field on and off, respectively. 
These rates are proportional to the spin-dependent absorption coefficients and in general\,\cite{Coutts2026},
\begin{align}
    \frac{\kappa_{\text{abs}}^{(j)}}{2\pi} = \frac{c}{n_{\text{g}}}\alpha^{(j)}\,,
    \label{eq:alpha2kappaMain}
\end{align}
where $c$ is the speed of light in vacuum. $n_{\text{g}}$ is the group index of the WGM, and $j = \text{(on)}, \text{(off)}$.
We define the average energy decay rate to be $\kappa = \kappa_{\text{ex}} + \kappa_{\text{i}} + \kappa_{\text{p}} + \kappa_{\text{abs}}^{(\text{off})} +  \Delta\kappa/2$, where $\Delta\kappa = \kappa_{\text{abs}}^{(\text{on})}-\kappa_{\text{abs}}^{(\text{off})}$ is the change in loss.
For simplicity, we set $\tau = 1$. Therefore, the change in transmission on resonance ($\Delta\omega_{\text{s}}=0$) is given by 
\begin{align}
    \Delta T &= \left|T^{(\text{on})} - T^{(\text{off})}\right| \\
    &= \frac{-16\,\Delta\kappa\,\kappa_{\text{ex}}\left[ \Delta\kappa^2 -4\,\kappa\,(\kappa-2\,\kappa_{\text{ex}}) \right]}{\left(\Delta\kappa^2 - 4\,\kappa^2\right)^2}.
\end{align}
Defining the contrast as we did before, 
\begin{align}
    C &= \frac{\Delta T}{T_0}\\
    &=\frac{16\,\Delta\kappa\,\kappa_{\text{ex}}\,\kappa^2\left[  4\,\kappa\,(\kappa-2\,\kappa_{\text{ex}}) - \Delta\kappa^2 \right]}{\left(\Delta\kappa^2 - 4\,\kappa^2\right)^2\left(\kappa-2\,\kappa_{\text{ex}}\right)^2},
\end{align}
where $T_0 = (1 - 2\,\kappa_{\text{ex}}/\kappa)^2$. 
Furthermore, to achieve the required intracavity intensity, the input power, and therefore the detected power, will depend on the cavity parameters. Specifically, the average cavity intensity on resonance can be expressed as~\cite{Coutts2026}
\begin{align}
    I_{\text{s,cav}} = \frac{2\,P_{\text{s}}\,c\,\kappa_{\text{ex}}}{n_{\text{g}}V_{\text{eff}}\,\kappa^2},
\end{align}
where $P_{\text{s}}$ is the power in the waveguide at the cavity and $V_{\text{eff}} = A\,L$, where $L$ is the effective cavity length and $A$ is the effective cavity cross-sectional area. 
Therefore, the optimal detected power can be expressed as 
\begin{align}
    P_{\text{det}} &= T_0\,P_{\text{s}}\\
    &= \frac{(\kappa-2\,\kappa_{\text{ex}})^2\,I_{\text{s,opt}}\,n_{\text{g}}\,A\,L}{2\,c\,\kappa_{\text{ex}}}\,.
\end{align}
Then, we can express the cw sensitivity of the side-coupled cavity as
\begin{align}
\begin{aligned}
    \eta_{\text{cav}} =& \beta\frac{\nu}{\sqrt{I_{\text{s,opt}}\,A}} \sqrt{\frac{c\,\kappa_{\text{ex}}}{L\,n_{\text{g}}}} 
    \\&\times\frac{\sqrt{2}\left(\Delta\kappa^2 - 4\,\kappa^2\right)^2\left(\kappa-2\,\kappa_{\text{ex}}\right)}{16\,\Delta\kappa\,\kappa_{\text{ex}}\,\kappa^2\left[ 4\,\kappa\,(\kappa-2\,\kappa_{\text{ex}}) - \Delta\kappa^2 \right]}.
    \label{eq:etaCav}
\end{aligned}
\end{align}
In our case, $\Delta\kappa\ll\kappa$, allowing us to simplify \Cref{eq:etaCav}:
\begin{align}
    \eta_{\text{cav}} &\approx \frac{\beta}{2\sqrt{2}} \frac{\nu}{\sqrt{I_{\text{s,opt}}\,A}}\sqrt{\frac{c}{L\,n_{\text{g}}}} \frac{\kappa}{\Delta\kappa\,\sqrt{\kappa_{\text{ex}}}}.
    \label{eq:etaCavSimple}
\end{align}
A useful proportionality can be expressed using \Cref{eq:absCoefficient,eq:etaCavSimple,eq:alpha2kappaMain}:
\begin{align}
    \eta_{\text{cav}} \propto \frac{1}{\,Q\sqrt{A\,L}},
\end{align}
where $Q = \omega_{\text{ZPL}}/\kappa$ is the total quality factor of the cavity.
In the case of an under-coupled cavity, $Q$ can be approximated by $Q_{\text{cav}} = \omega_{\text{ZPL}}/(\kappa_{\text{i}} + \kappa_{\text{p}} + \kappa_{\text{abs}}^{(\text{off})} +  \Delta\kappa/2 )$.

For an absorption-limited cavity $\left(\kappa \approx \kappa_{\text{ex}} + \kappa_{\text{abs}}^{(\text{off})} + \Delta\kappa/2\right)$, we find that \Cref{eq:etaCavSimple} is minimized near critical coupling ($ \kappa_{\text{ex}} \approx  \kappa_{\text{abs}}^{(\text{off})} + \Delta\kappa/2$). 
In practice, operation near critical coupling suppresses the detected optical power, making detector efficiency and technical noise increasingly important. 
We therefore interpret the critical-coupling result as an ideal lower bound on the sensitivity.

We can then rewrite \Cref{eq:etaCavSimple} as a function of the NV singlet-absorption coefficients:
\begin{align}
    \eta_{\text{cav}} \approx \beta\frac{\nu}{\sqrt{I_{\text{s,opt}}\,A}}\frac{\sqrt{\alpha_0}}{\Delta\alpha\sqrt{L}}.\label{eq:etaCavSimp}
\end{align}

Comparing \Cref{eq:etaFPcavSimp,eq:etaCavSimp} we find that, in the small-sensor limit, the side-coupled cavity achieves approximately the same improvement as the etalon, assuming near-optimal coupling.

To predict the PSNL sensitivity of our experimental cavities, we extract the cavity parameters by fitting the transmission spectra found in \Cref{fig:fig2_ODMR}~(c) and \Cref{fig:fig11_highQ}~(a). Each set of parameters is then used alongside \Cref{eq:etaCav} to calculate the sensitivities presented in \Cref{tab:tab3_cavityParameters}, which are scaled by the fibre-taper transmission and detection efficiencies to better compare to experiment.

\bibliography{bibliographies/main}

\end{document}